\documentclass[aps,prd,reprint,superscriptaddress]{revtex4-1}
\usepackage{hyperref}
\usepackage{graphicx}
\usepackage{enumerate}
\hypersetup{
    pdfnewwindow=true,      
    colorlinks=true,        
    linkcolor=blue,         
    citecolor=black,        
    filecolor=blue,         
    urlcolor=blue           
}

\usepackage{subfigure}
\usepackage{amsmath}
\usepackage{amssymb}
\usepackage{amsfonts}
\RequirePackage{lineno}
\usepackage{color}

\def\be{\begin{equation}}
\def\ee{\end{equation}}

\providecommand{\apj}[0]{Astrophys. J.}
\providecommand{\apjl}[0]{Astrophys. J. Lett.}
\providecommand{\apjs}[0]{Astrophys. J. Supp. Ser. }

\providecommand{\aap}[0]{Astron. Astrophys. }

\providecommand{\araa}[0]{Ann.\ Rev. Astron. Astroph. }

\providecommand{\mnras}[0]{Mon. Not. Roy. Astron. Soc. }

\providecommand{\prl}[0]{Phys. Rev. Lett.}
\providecommand{\prd}{Phys. Rev. D.}

\providecommand{\apss}[0]{Astrophys. Space Sci.}

\providecommand{\jcap}[0]{JCAP}

\begin{document}

\title{Multimessenger Search for Sources of Gravitational Waves and High-Energy Neutrinos: Results for Initial LIGO-Virgo and IceCube}

\affiliation{III. Physikalisches Institut, RWTH Aachen University, D-52056 Aachen, Germany}
\affiliation{School of Chemistry \& Physics, University of Adelaide, Adelaide SA, 5005 Australia}
\affiliation{Dept.~of Physics and Astronomy, University of Alaska Anchorage, 3211 Providence Dr., Anchorage, AK 99508, USA}
\affiliation{CTSPS, Clark-Atlanta University, Atlanta, GA 30314, USA}
\affiliation{School of Physics and Center for Relativistic Astrophysics, Georgia Institute of Technology, Atlanta, GA 30332, USA}
\affiliation{Dept.~of Physics, Southern University, Baton Rouge, LA 70813, USA}
\affiliation{Dept.~of Physics, University of California, Berkeley, CA 94720, USA}
\affiliation{Lawrence Berkeley National Laboratory, Berkeley, CA 94720, USA}
\affiliation{Institut f\"ur Physik, Humboldt-Universit\"at zu Berlin, D-12489 Berlin, Germany}
\affiliation{Fakult\"at f\"ur Physik \& Astronomie, Ruhr-Universit\"at Bochum, D-44780 Bochum, Germany}
\affiliation{Physikalisches Institut, Universit\"at Bonn, Nussallee 12, D-53115 Bonn, Germany}
\affiliation{Universit\'e Libre de Bruxelles, Science Faculty CP230, B-1050 Brussels, Belgium}
\affiliation{Vrije Universiteit Brussel, Dienst ELEM, B-1050 Brussels, Belgium}
\affiliation{Dept.~of Physics, Chiba University, Chiba 263-8522, Japan}
\affiliation{Dept.~of Physics and Astronomy, University of Canterbury, Private Bag 4800, Christchurch, New Zealand}
\affiliation{Dept.~of Physics, University of Maryland, College Park, MD 20742, USA}
\affiliation{Dept.~of Physics and Center for Cosmology and Astro-Particle Physics, Ohio State University, Columbus, OH 43210, USA}
\affiliation{Dept.~of Astronomy, Ohio State University, Columbus, OH 43210, USA}
\affiliation{Niels Bohr Institute, University of Copenhagen, DK-2100 Copenhagen, Denmark}
\affiliation{Dept.~of Physics, TU Dortmund University, D-44221 Dortmund, Germany}
\affiliation{Dept.~of Physics, University of Alberta, Edmonton, Alberta, Canada T6G 2E1}
\affiliation{Erlangen Centre for Astroparticle Physics, Friedrich-Alexander-Universit\"at Erlangen-N\"urnberg, D-91058 Erlangen, Germany}
\affiliation{D\'epartement de physique nucl\'eaire et corpusculaire, Universit\'e de Gen\`eve, CH-1211 Gen\`eve, Switzerland}
\affiliation{Dept.~of Physics and Astronomy, University of Gent, B-9000 Gent, Belgium}
\affiliation{Dept.~of Physics and Astronomy, University of California, Irvine, CA 92697, USA}
\affiliation{Dept.~of Physics and Astronomy, University of Kansas, Lawrence, KS 66045, USA}
\affiliation{Dept.~of Astronomy, University of Wisconsin, Madison, WI 53706, USA}
\affiliation{Dept.~of Physics and Wisconsin IceCube Particle Astrophysics Center, University of Wisconsin, Madison, WI 53706, USA}
\affiliation{Institute of Physics, University of Mainz, Staudinger Weg 7, D-55099 Mainz, Germany}
\affiliation{Universit\'e de Mons, 7000 Mons, Belgium}
\affiliation{Technische Universit\"at M\"unchen, D-85748 Garching, Germany}
\affiliation{Bartol Research Institute and Dept.~of Physics and Astronomy, University of Delaware, Newark, DE 19716, USA}
\affiliation{Dept.~of Physics, University of Oxford, 1 Keble Road, Oxford OX1 3NP, UK}
\affiliation{Physics Department, South Dakota School of Mines and Technology, Rapid City, SD 57701, USA}
\affiliation{Dept.~of Physics, University of Wisconsin, River Falls, WI 54022, USA}
\affiliation{Oskar Klein Centre and Dept.~of Physics, Stockholm University, SE-10691 Stockholm, Sweden}
\affiliation{Dept.~of Physics and Astronomy, Stony Brook University, Stony Brook, NY 11794-3800, USA}
\affiliation{Dept.~of Physics, Sungkyunkwan University, Suwon 440-746, Korea}
\affiliation{Dept.~of Physics, University of Toronto, Toronto, Ontario, Canada, M5S 1A7}
\affiliation{Dept.~of Physics and Astronomy, University of Alabama, Tuscaloosa, AL 35487, USA}
\affiliation{Dept.~of Astronomy and Astrophysics, Pennsylvania State University, University Park, PA 16802, USA}
\affiliation{Dept.~of Physics, Pennsylvania State University, University Park, PA 16802, USA}
\affiliation{Dept.~of Physics and Astronomy, Uppsala University, Box 516, S-75120 Uppsala, Sweden}
\affiliation{Dept.~of Physics, University of Wuppertal, D-42119 Wuppertal, Germany}
\affiliation{DESY, D-15735 Zeuthen, Germany}

\author{M.~G.~Aartsen}
\affiliation{School of Chemistry \& Physics, University of Adelaide, Adelaide SA, 5005 Australia}
\author{M.~Ackermann}
\affiliation{DESY, D-15735 Zeuthen, Germany}
\author{J.~Adams}
\affiliation{Dept.~of Physics and Astronomy, University of Canterbury, Private Bag 4800, Christchurch, New Zealand}
\author{J.~A.~Aguilar}
\affiliation{D\'epartement de physique nucl\'eaire et corpusculaire, Universit\'e de Gen\`eve, CH-1211 Gen\`eve, Switzerland}
\author{M.~Ahlers}
\affiliation{Dept.~of Physics and Wisconsin IceCube Particle Astrophysics Center, University of Wisconsin, Madison, WI 53706, USA}
\author{M.~Ahrens}
\affiliation{Oskar Klein Centre and Dept.~of Physics, Stockholm University, SE-10691 Stockholm, Sweden}
\author{D.~Altmann}
\affiliation{Erlangen Centre for Astroparticle Physics, Friedrich-Alexander-Universit\"at Erlangen-N\"urnberg, D-91058 Erlangen, Germany}
\author{T.~Anderson}
\affiliation{Dept.~of Physics, Pennsylvania State University, University Park, PA 16802, USA}
\author{C.~Arguelles}
\affiliation{Dept.~of Physics and Wisconsin IceCube Particle Astrophysics Center, University of Wisconsin, Madison, WI 53706, USA}
\author{T.~C.~Arlen}
\affiliation{Dept.~of Physics, Pennsylvania State University, University Park, PA 16802, USA}
% ...
\collaboration{IceCube Collaboration}
\noaffiliation

\author{%
J.~Aasi,$^{46}$
B.~P.~Abbott,$^{46}$
R.~Abbott,$^{46}$
T.~Abbott,$^{47}$
M.~R.~Abernathy,$^{46}$
F.~Acernese,$^{48,49}$
K.~Ackley,$^{50}$
C.~Adams,$^{51}$
T.~Adams,$^{52}$
P.~Addesso,$^{53}$
% ...
}\noaffiliation

\affiliation {LIGO, California Institute of Technology, Pasadena, CA 91125, USA }
\affiliation {Louisiana State University, Baton Rouge, LA 70803, USA }
\affiliation {Universit\`a di Salerno, Fisciano, I-84084 Salerno, Italy }
\affiliation {INFN, Sezione di Napoli, Complesso Universitario di Monte S.Angelo, I-80126 Napoli, Italy }
\affiliation {University of Florida, Gainesville, FL 32611, USA }
\affiliation {LIGO Livingston Observatory, Livingston, LA 70754, USA }
\affiliation {Cardiff University, Cardiff, CF24 3AA, United Kingdom }
\affiliation {University of Sannio at Benevento, I-82100 Benevento, Italy, and INFN, Sezione di Napoli, I-80100 Napoli, Italy. }
\affiliation {Albert-Einstein-Institut, Max-Planck-Institut f\"ur Gravitationsphysik, D-30167 Hannover, Germany }
\affiliation {Nikhef, Science Park, 1098 XG Amsterdam, The Netherlands }
\affiliation {LIGO, Massachusetts Institute of Technology, Cambridge, MA 02139, USA }
\affiliation {Instituto Nacional de Pesquisas Espaciais, 12227-010 - S\~{a}o Jos\'{e} dos Campos, SP, Brazil }
\affiliation {International Centre for Theoretical Sciences, Tata Institute of Fundamental Research, Bangalore 560012, India. }
\affiliation {Syracuse University, Syracuse, NY 13244, USA }
\affiliation {University of Wisconsin--Milwaukee, Milwaukee, WI 53201, USA }
\affiliation {Leibniz Universit\"at Hannover, D-30167 Hannover, Germany }
\affiliation {Universit\`a di Siena, I-53100 Siena, Italy }
\affiliation {INFN, Sezione di Pisa, I-56127 Pisa, Italy }
\affiliation {Stanford University, Stanford, CA 94305, USA }
\affiliation {The University of Mississippi, University, MS 38677, USA }
\affiliation {California State University Fullerton, Fullerton, CA 92831, USA }
\affiliation {INFN, Sezione di Roma, I-00185 Roma, Italy }
\affiliation {University of Birmingham, Birmingham, B15 2TT, United Kingdom }
\affiliation {Albert-Einstein-Institut, Max-Planck-Institut f\"ur Gravitationsphysik, D-14476 Golm, Germany }
\affiliation {Montana State University, Bozeman, MT 59717, USA }
\affiliation {European Gravitational Observatory (EGO), I-56021 Cascina, Pisa, Italy }
\affiliation {LIGO Hanford Observatory, Richland, WA 99352, USA }
\affiliation {SUPA, University of Glasgow, Glasgow, G12 8QQ, United Kingdom }
\affiliation {APC, AstroParticule et Cosmologie, Universit\'e Paris Diderot, CNRS/IN2P3, CEA/Irfu, Observatoire de Paris, Sorbonne Paris Cit\'e, 10, rue Alice Domon et L\'eonie Duquet, F-75205 Paris Cedex 13, France }
\affiliation {Columbia University, New York, NY 10027, USA }
\affiliation {Universit\`a di Pisa, I-56127 Pisa, Italy }
\affiliation {CAMK-PAN, 00-716 Warsaw, Poland }
\affiliation {Astronomical Observatory Warsaw University, 00-478 Warsaw, Poland }
\affiliation {Universit\`a degli Studi di Genova, I-16146 Genova, Italy }
\affiliation {INFN, Sezione di Genova, I-16146 Genova, Italy }
\affiliation {Faculty of Physics, Lomonosov Moscow State University, Moscow 119991, Russia }
\affiliation {LAL, Universit\'e Paris-Sud, IN2P3/CNRS, F-91898 Orsay, France }
\affiliation {NASA/Goddard Space Flight Center, Greenbelt, MD 20771, USA }
\affiliation {University of Western Australia, Crawley, WA 6009, Australia }
\affiliation {Department of Astrophysics/IMAPP, Radboud University Nijmegen, P.O. Box 9010, 6500 GL Nijmegen, The Netherlands }
\affiliation {Universit\'e Nice-Sophia-Antipolis, CNRS, Observatoire de la C\^ote d'Azur, F-06304 Nice, France }
\affiliation {MTA Atomki, Debrecen, Hungary }
\affiliation {Institut de Physique de Rennes, CNRS, Universit\'e de Rennes 1, F-35042 Rennes, France }
\affiliation {Laboratoire d'Annecy-le-Vieux de Physique des Particules (LAPP), Universit\'e de Savoie, CNRS/IN2P3, F-74941 Annecy-le-Vieux, France }
\affiliation {Washington State University, Pullman, WA 99164, USA }
\affiliation {Inter-University Centre for Astronomy and Astrophysics, Pune - 411007, India }
\affiliation {INFN, Sezione di Perugia, I-06123 Perugia, Italy }
\affiliation {Yukawa Institute for Theoretical Physics, Kyoto University, Kyoto 606-8502, Japan }
\affiliation {Universit\`a degli Studi di Urbino 'Carlo Bo', I-61029 Urbino, Italy }
\affiliation {INFN, Sezione di Firenze, I-50019 Sesto Fiorentino, Firenze, Italy }
\affiliation {University of Oregon, Eugene, OR 97403, USA }
\affiliation {Laboratoire Kastler Brossel, ENS, CNRS, UPMC, Universit\'e Pierre et Marie Curie, F-75005 Paris, France }
\affiliation {VU University Amsterdam, 1081 HV Amsterdam, The Netherlands }
\affiliation {University of Maryland, College Park, MD 20742, USA }
\affiliation {University of Massachusetts Amherst, Amherst, MA 01003, USA }
\affiliation {Laboratoire des Mat\'eriaux Avanc\'es (LMA), IN2P3/CNRS, Universit\'e de Lyon, F-69622 Villeurbanne, Lyon, France }
\affiliation {Universitat de les Illes Balears, E-07122 Palma de Mallorca, Spain }
\affiliation {Universit\`a di Napoli 'Federico II', Complesso Universitario di Monte S.Angelo, I-80126 Napoli, Italy }
\affiliation {Canadian Institute for Theoretical Astrophysics, University of Toronto, Toronto, Ontario, M5S 3H8, Canada }
\affiliation {Tsinghua University, Beijing 100084, China }
\affiliation {University of Michigan, Ann Arbor, MI 48109, USA }
\affiliation {INFN, Sezione di Roma Tor Vergata, I-00133 Roma, Italy }
\affiliation {National Tsing Hua University, Hsinchu Taiwan 300 }
\affiliation {Charles Sturt University, Wagga Wagga, NSW 2678, Australia }
\affiliation {Caltech-CaRT, Pasadena, CA 91125, USA }
\affiliation {Pusan National University, Busan 609-735, Korea }
\affiliation {Australian National University, Canberra, ACT 0200, Australia }
\affiliation {Carleton College, Northfield, MN 55057, USA }
\affiliation {Universit\`a di Roma Tor Vergata, I-00133 Roma, Italy }
\affiliation {INFN, Gran Sasso Science Institute, I-67100 L'Aquila, Italy }
\affiliation {Universit\`a di Roma 'La Sapienza', I-00185 Roma, Italy }
\affiliation {University of Brussels, Brussels 1050 Belgium }
\affiliation {Sonoma State University, Rohnert Park, CA 94928, USA }
\affiliation {The George Washington University, Washington, DC 20052, USA }
\affiliation {University of Cambridge, Cambridge, CB2 1TN, United Kingdom }
\affiliation {University of Minnesota, Minneapolis, MN 55455, USA }
\affiliation {The University of Texas at Brownsville, Brownsville, TX 78520, USA }
\affiliation {The University of Sheffield, Sheffield S10 2TN, United Kingdom }
\affiliation {Wigner RCP, RMKI, H-1121 Budapest, Konkoly Thege Mikl\'os \'ut 29-33, Hungary }
\affiliation {Argentinian Gravitational Wave Group, Cordoba Cordoba 5000, Argentina }
\affiliation {Universit\`a di Trento, I-38050 Povo, Trento, Italy }
\affiliation {INFN, Gruppo Collegato di Trento, I-38050 Povo, Trento, Italy }
\affiliation {Northwestern University, Evanston, IL 60208, USA }
\affiliation {Montclair State University, Montclair, NJ 07043, USA }
\affiliation {The Pennsylvania State University, University Park, PA 16802, USA }
\affiliation {MTA E\"otv\"os University, `Lendulet' A. R. G., Budapest 1117, Hungary }
\affiliation {Universit\`a di Perugia, I-06123 Perugia, Italy }
\affiliation {Rutherford Appleton Laboratory, HSIC, Chilton, Didcot, Oxon, OX11 0QX, United Kingdom }
\affiliation {Embry-Riddle Aeronautical University, Prescott, AZ 86301, USA }
\affiliation {Seoul National University, Seoul 151-742, Korea }
\affiliation {Perimeter Institute for Theoretical Physics, Waterloo, Ontario, N2L 2Y5, Canada }
\affiliation {American University, Washington, DC 20016, USA }
\affiliation {College of William and Mary, Williamsburg, VA 23187, USA }
\affiliation {University of Adelaide, Adelaide, SA 5005, Australia }
\affiliation {Raman Research Institute, Bangalore, Karnataka 560080, India }
\affiliation {Korea Institute of Science and Technology Information, Daejeon 305-806, Korea }
\affiliation {Bia{\l }ystok University, 15-424 Bia{\l }ystok, Poland }
\affiliation {University of Southampton, Southampton, SO17 1BJ, United Kingdom }
\affiliation {IISER-TVM, CET Campus, Trivandrum Kerala 695016, India }
\affiliation {Institute of Applied Physics, Nizhny Novgorod, 603950, Russia }
\affiliation {Hanyang University, Seoul 133-791, Korea }
\affiliation {NCBJ, 05-400 \'Swierk-Otwock, Poland }
\affiliation {IM-PAN, 00-956 Warsaw, Poland }
\affiliation {Institute for Plasma Research, Bhat, Gandhinagar 382428, India }
\affiliation {The University of Melbourne, Parkville, VIC 3010, Australia }
\affiliation {INFN, Sezione di Padova, I-35131 Padova, Italy }
\affiliation {Monash University, Victoria 3800, Australia }
\affiliation {SUPA, University of Strathclyde, Glasgow, G1 1XQ, United Kingdom }
\affiliation {Louisiana Tech University, Ruston, LA 71272, USA }
\affiliation {ESPCI, CNRS, F-75005 Paris, France }
\affiliation {Universit\`a di Camerino, Dipartimento di Fisica, I-62032 Camerino, Italy }
\affiliation {Southern University and A\&M College, Baton Rouge, LA 70813, USA }
\affiliation {IISER-Kolkata, Mohanpur, West Bengal 741252, India }
\affiliation {National Institute for Mathematical Sciences, Daejeon 305-390, Korea }
\affiliation {Hobart and William Smith Colleges, Geneva, NY 14456, USA }
\affiliation {RRCAT, Indore MP 452013, India }
\affiliation {Tata Institute for Fundamental Research, Mumbai 400005, India }
\affiliation {SUPA, University of the West of Scotland, Paisley, PA1 2BE, United Kingdom }
\affiliation {Institute of Astronomy, 65-265 Zielona G\'ora, Poland }
\affiliation {Indian Institute of Technology, Gandhinagar Ahmedabad Gujarat 382424, India }
\affiliation {Instituto de F\'\i sica Te\'orica, Univ. Estadual Paulista/ICTP South American Institute for Fundamental Research, S\~ao Paulo SP 01140-070, Brazil }
\affiliation {Andrews University, Berrien Springs, MI 49104, USA }
\affiliation {Trinity University, San Antonio, TX 78212, USA }
\affiliation {University of Washington, Seattle, WA 98195, USA }
\affiliation {Rochester Institute of Technology, Rochester, NY 14623, USA }
\affiliation {Southeastern Louisiana University, Hammond, LA 70402, USA }
\affiliation {Abilene Christian University, Abilene, TX 79699, USA }
\collaboration{The LIGO Scientific Collaboration and the Virgo Collaboration}
\noaffiliation

\begin{abstract}
We report the results of a multimessenger search for coincident signals from the LIGO and Virgo gravitational-wave observatories and the partially completed IceCube high-energy neutrino detector, including periods of joint operation between 2007-2010. These include parts of the 2005-2007 run and the 2009-2010 run for LIGO-Virgo, and IceCube's observation periods with 22, 59 and 79 strings. We find no significant coincident events, and use the search results to derive upper limits on the rate of joint sources for a range of source emission parameters. For the optimistic assumption of gravitational-wave emission energy of $10^{-2}$\,M$_\odot$c$^2$ at $\sim 150$\,Hz with $\sim 60$\,ms duration, and high-energy neutrino emission of $10^{51}$\,erg comparable to the isotropic gamma-ray energy of gamma-ray bursts, we limit the source rate below $1.6 \times 10^{-2}$\,Mpc$^{-3}$yr$^{-1}$. We also examine how combining information from gravitational waves and neutrinos will aid discovery in the advanced gravitational-wave detector era.
\end{abstract}

\maketitle

\section{Introduction}

Gravitational-wave (GW) and high-energy-neutrino observations represent important novel opportunities in studying a variety of cosmic processes, providing complementary information to electromagnetic searches. They carry information on the origin of high-energy emission as well as the dynamics and structure of the process driving the emission \cite{LIGO0034-4885-72-7-076901,2003PhRvD..68h3001R,PhysRevD.86.083007,2012Natur.484..351A,2013PhRvL.110x1101B,PhysRevLett.108.231101,2012ApJ...752...29H}, which may be difficult to decipher via electromagnetic observations (\cite{2013RvMP...85.1401A,2013CQGra..30l3001B} and references therein).

GWs are produced by the bulk motion of the progenitor, typically carrying information on the dynamics of the source's central region. High-energy neutrinos, on the other hand, require hadron acceleration in, e.g., relativistic outflows from a central engine. Astrophysical processes that produce GWs may also drive relativistic outflows, which can emit high-energy radiation, such as GeV-PeV neutrinos or gamma rays. Detecting GWs along with other forms of radiation from a common source would shed light on the internal processes within the source, and could increase detection confidence.

The search for common sources of GWs and high-energy-neutrino has recently become possible with the construction and upgrade of large-scale observatories. GW detectors include LIGO \cite{LIGO0034-4885-72-7-076901} and Virgo \cite{0264-9381-28-11-114002}, which are being upgraded to second-generation detectors \cite{2010CQGra..27h4006H,AdV}, and GEO \cite{2010CQGra..27h4003G}, which is currently acquiring data. Another advanced GW detector, KAGRA \cite{2012CQGra..29l4007S}, is being constructed in Japan. The possibility to construct a third advanced LIGO observational facility in India is also currently being explored \cite{LIGOindia}. With the construction of the advanced LIGO detectors finishing in 2015, multiple detections of GWs from compact binary mergers are expected after 2016-2018 \cite{2013arXiv1304.0670L}. High-energy neutrino observatories currently in operation include IceCube \cite{2006APh....26..155I,2009NIMPA.601..294A,2010NIMPA.618..139A} -- a cubic-kilometer detector at the geographic South Pole -- and \textsc{Antares} \cite{ANTARES} in the Mediterranean sea. \textsc{Antares} is planned to be followed by a multi-cubic-kilometer detector in the Mediterranean sea called KM3NeT in the following years \cite{deJong2010445}. The Baikal Neutrino Telescope, operating at Lake Baikal, is also planned to be upgraded to a km$^3$ volume \cite{Avrorin2011S13}. IceCube is also sensitive to low energy (MeV) thermal neutrinos from nearby supernovae \cite{2011A&A...535A.109A}, and contributes to the Supernova Early Warning System (SNEWS) network along with several other neutrino detectors, including Super-Kamiokande \cite{2003NIMPA.501..418F}, Borexino \cite{2002APh....16..361C}, the Large Volume Detector \cite{2008APh....28..516A} and the Sudbury Neutrino Observatory \cite{2000NIMPA.449..172B}.

High-energy neutrinos of cosmic origin have recently been observed, for the first time, by IceCube \cite{2013PhRvL.111b1103A,IceCubeScience}. Their detection represents a major step towards multimessenger astronomy. In an all-sky search for the most energetic ($> 30$\,TeV) neutrinos, a total of 28 astrophysical neutrino candidates were identified over a period of 2 years, significantly above the number of expected atmospheric background ($\approx 10$ events). The neutrinos have no clear directional or temporal clustering. While many sources of astrophysical origin have been suggested (see, e.g., \cite{2013JCAP...06..030C,2013PhRvD..88h3007W,2013PhRvL.111d1103K,2013JCAP...01..028R,2013arXiv1305.7404S,2013PhRvD..87f3011H,2013PhRvD..88l1301M,2014PhRvD..89h3003A,2013arXiv1310.7194G,2013PhRvD..88h1302R}, so far there is no compelling evidence for a specific source population.

Several classes of astrophysical transients that are detected electromagnetically are also associated with significant GW and high-energy neutrino emission. These processes include the gamma-ray bursts \cite{2012PhRvL.108w1101H,1997PhRvL..78.2292W,2005PhRvL..95f1103A,2006ApJ...651L...5M,2003ApJ...589..861K,2002ApJ...579L..63D,2004PhRvD..69d4007V,2007ApJ...658.1173P,2011PhRvL.106y1102K,2009ApJ...702.1171C,2013RvMP...85.1401A,2013RvMP...85.1401A}, core-collapse supernovae with rapidly rotating cores \cite{2011LRR....14....1F,2006RPPh...69..971K,2005PhRvD..71b4014S}, flares from soft gamma repeaters \cite{2005ApJ...633.1013I,2012PhRvD..85b4030Z,2013PhRvD..87j3008M}, or even cosmic string cusps \cite{2000PhRvL..85.3761D,2011PhRvD..84h5006B,2006PhRvD..73j5001S,2014PhRvL.112m1101A}.

Gamma-ray bursts (GRBs) are one of the most promising sources for joint GW+neutrino observations. GRBs are intense flashes of $\sim$\,MeV photons of typically extragalactic origin, occurring a few times a day in the observable universe \cite{2013APh....43..134M}. They are thought to be produced in relativistic outflows likely driven by accretion onto a compact object \cite{1993ApJ...405..273W,1999ApJ...524..262M}, or by a rapidly rotating neutron star \cite{1992Natur.357..472U}. Hadrons accelerated in the relativistic outflows will result in the production of high-energy neutrinos through interaction with photons or other hadrons \cite{2005PhRvL..95f1103A,2006ApJ...651L...5M}. The central engines that drive the outflow can be formed by (i) the collapse of the rapidly rotating core of a massive star, which can also produce supernovae \cite{2006ARA&A..44..507W}, or by (ii) the merger of two neutron stars or a neutron star and a black hole \cite{1989Natur.340..126E}.  Stellar core-collapse with rapidly rotating cores are thought to be the origin of long ($\gtrsim2$\,s) GRBs. They can lead to GW emission via, e.g., rotational instabilities in the newly formed protoneutron stars, non-axisymmetric instabilities of the accretion disk, or the fragmentation of the collapsing core \cite{1997A&A...317..140M,2009CQGra..26f3001O,2013CRPhy..14..318K}. Compact binary neutron star or neutron-star-black-hole mergers are thought to be the progenitors of short ($\lesssim2$\,s) GRBs \cite{2013ApJ...778L..16H,2011ApJ...732L...6R}. They emit strong GWs as the two compact objects inspiral and merge, in the sensitive frequency band of LIGO/Virgo \cite{2009LRR....12....2S,1999CQGra..16R...1R,2005PhRvD..71h4021S}.

Another promising source type, soft-gamma repeaters (SGR), are sources of short bursts of gamma rays emitted at irregular intervals. They occasionally emit giant gamma-ray/X-ray flares that can be detected from the Milky Way or nearby galaxies within a few megaparsecs, at a rate lower than the GRB rate (e.g., \cite{2010MNRAS.403..342H}). SGRs are thought to be highly magnetized neutron stars. Their gamma outbursts may be a consequence of the tectonic activity of the neutron star crust (starquake). The seismic vibration of a neutron star following a starquake gives rise to GW emission (e.g., \cite{2005PhRvL..95u1101O,2006cosp...36..168W,2013PhRvD..87j3008M}). Tectonic activity in the neutron star is followed by the reconfiguration of the neutron star's magnetic fields, resulting in X-ray and gamma-ray radiation \cite{1992ApJ...392L...9D,1995MNRAS.275..255T}. Sudden magnetic reconfiguration may also accelerate protons and other nuclei, leading to the production of high-energy neutrinos \cite{2005ApJ...634L..89G,2005ApJ...633.1013I}.

For joint GW+neutrino detection, a particularly interesting subset of these sources are those that have faint electromagnetic emission, such as choked GRBs \cite{2001PhRvL..87q1102M,2007arXiv0711.4791K,2008PhRvD..77f3007H} for which the relativistic outflow stalls before it can break out of the stelar envelope, and low-luminosity GRBs with mildly relativistic ejecta \cite{2006ApJ...651L...5M,2007PhRvD..76h3009W}.

The coincident detectability of GWs and high-energy neutrinos, and the scientific potential of such searches, was first recognized in \cite{2008CQGra..25k4039A}, and later independently in \cite{2009NIMPA.602..268P}. The population of cosmic events producing both GWs and high-energy neutrino emission was first constrained observationally \cite{PhysRevLett.107.251101} with data from the initial LIGO-Virgo detectors and the partially completed IceCube detector (see Section \ref{section:rateUL}). While the population of cosmic events expected to produce detectable joint emission falls below these early observational limits, the results were used to estimate the projected upper limits achievable with advanced GW detectors and the completed IceCube, which can provide meaningful constraints or detection.

A search for coincident GWs and high-energy neutrinos has recently been carried out using the initial LIGO-Virgo observatories and the partially completed {\sc Antares} detector \cite{2013JCAP...06..008A}. The search found no significant coincident events. {\sc Antares} is mainly sensitive to TeV neutrino sources from the southern hemisphere, making this a complementary search to that performed with IceCube, which is mainly sensitive to the northern hemisphere at TeV energies.

Recognizing the scientific potential of multimessenger observations, various initiatives aim to use GWs and neutrinos in combination with electromagnetic observations. For instance, multiple astrophysical event candidates from both the initial LIGO-Virgo detectors \cite{2007AAS...211.9903P,2008CQGra..25r4034K} and IceCube \cite{2012A&A...539A..60A,2007APh....27..533K} have been followed up by EM telescopes: QUEST, TAROT, ZADKO, Pi of the sky, ROTSE, SkyMapper, the Palomar Transient Factory, and the Swift satellite \cite{2012A&A...539A.124L,2012A&A...541A.155A,2012ApJS..203...28E,2013arXiv1310.2314T}. Sub-threshold GRB candidates detected through gamma-rays are also being used for searches for coincident GW signals \cite{subthresholdGRB,subthresholdGRB2,2013arXiv1303.2174B}. The SNEWS network \cite{2004NJPh....6..114A} aims to use sub-threshold MeV neutrino signals from neutrino observatories around the world to trigger electromagnetic follow-up searches for supernovae. The network utilizes the fact that neutrinos arrive up to several hours earlier than the detectable electromagnetic signal from supernovae, as was the case for the first confirmed astronomical MeV neutrino source, SN\,1987A \cite{PhysRevLett.58.1490,PhysRevLett.58.1494}. A current initiative for multimessenger searches in the weak-signal limit is the proposed Astrophysical Multimessenger Observatory Network (AMON; \cite{2013APh....45...56S}). AMON is planned to perform a real-time correlation analysis of sub-threshold signals from available astronomical messengers; photons, low and high-energy neutrinos, cosmic rays, and GWs. These sub-threshold triggers will be used to (1) increase the detectable number of sources and (2) trigger follow-up searches that can further enhance the significance of a potential coincident detection.

In this paper we present a multimessenger search for joint sources of GWs and high-energy neutrinos using the initial LIGO and Virgo GW detectors together with the partially complete IceCube detector in its 22, 59 and 79-string configurations. The analysis is based on the method described in \cite{2012PhRvD..85j3004B}. In Section \ref{section:multianalysis} we briefly review the search method and data-analysis details. In Section \ref{section:results} we present the results.  Observing no evidence of a GW+neutrino event we set upper limits on the rate of GW+neutrino sources as a function of the source energy budget.
We summarize our findings and discuss the outlook for future observations in Section \ref{section:discussion}.

\section{Multimessenger analysis}
\label{section:multianalysis}

In this section we present the status of GW and neutrino observations and the analysis we used to combine their results; a more detailed description of the method can be found in \cite{2012PhRvD..85j3004B}. Hereafter, we use the word ``neutrinos" to refer exclusively to high-energy neutrinos.

\subsection{Gravitational-wave data}
\label{section:GWdata}

LIGO \cite{LIGO0034-4885-72-7-076901} and Virgo \cite{0264-9381-28-11-114002} are km-scale interferometric GW observatories. LIGO has two Michelson interferometers in the U.S., both with 4-km arm length. One of the detectors is in Hanford, Washington (which we refer to as H1), while the other one is located in Livingston, Louisiana (which we refer to as L1). During the 2005-2007 observation period, an additional detector with 2-km arm length was operating at the Hanford site (which we refer to as H2). H2's location and orientation were identical to that of H1. The LIGO detectors were sensitive to GWs in the frequency band of $40-7000$\,Hz. Virgo is a Michelson interferometer with 3-km arm length (which we refer to as V1), located near Cascina, Italy. Virgo is sensitive to GWs in a frequency band similar to that of LIGO. For the present search, data from initial Virgo's three observation periods (VSR1-3) were used.

LIGO and Virgo detect GWs by monitoring the relative displacement of test masses (mirrors). An incoming GW changes the distance between test masses with an amount proportional to their distance. The detectors measure the fractional difference $h=(L_1-L_2) / L$ between the lengths of the two interferometer arms $L_1$ and $L_2$ ($L_2\simeq L_1=L$), which is referred to as strain. The two GW polarizations, denoted with $h_+$ and $h_\times$, independently contribute to the detector strain. The detector strain also depends on the relative orientation of the detector and the GW, which is described by so-called antenna response factors $F_+$ and $F_\times$: $h(t)=F_+ h_+(t) + F_\times h_\times (t)$ \cite{1992Sci...256..325A}.

The astrophysical reach of GW detectors can be characterized, e.g., by the so-called horizon distance: the maximum distance at which the GW signal from an optimally oriented and optimally located source can be detected, with detection defined as a single-detector signal-to-noise ratio of 8. It is easiest to establish this horizon distance for compact binary mergers for which the gravitational waveform is known to sufficient precision. This horizon distance reached tens of megaparsecs for the initial LIGO-Virgo detectors \cite{2010arXiv1003.2481T,2012arXiv1203.2674T}. The GW output, and therefore the horizon distances, of stellar core collapse with rapidly rotating cores and SGR starquakes are less certain \cite{2013CQGra..30l3001B}. Adopting an optimistic standard-siren GW emission with energy $E_{\rm gw} = 10^{-2}$\,M$_\odot$c$^2$ and frequency $f_0=150$\,Hz (e.g., \cite{2014arXiv1403.6639T}), such a GW signal could be detected out to $\gtrsim 10$\,Mpc with initial LIGO-Virgo. For SGRs, the GW counterpart of giant flares may be detectable, in the most optimistic scenarios, from within the Galaxy \cite{2008PhRvL.101u1102A}.

For the present search, data from the initial LIGO detectors' fifth (S5) and sixth (S6) observation periods were used. During the S5/VSR1 observation period, a small fraction of the data was taken with only the H1, L1 and V1 detectors, i.e. without H2. Given that this period was significantly shorter than when \{H1,H2,L1,V1\} were operational, and for simplicity, we exclude this period from the present analysis.

\begin{figure*}
\begin{center}
\resizebox{0.7\textwidth}{!}{\includegraphics{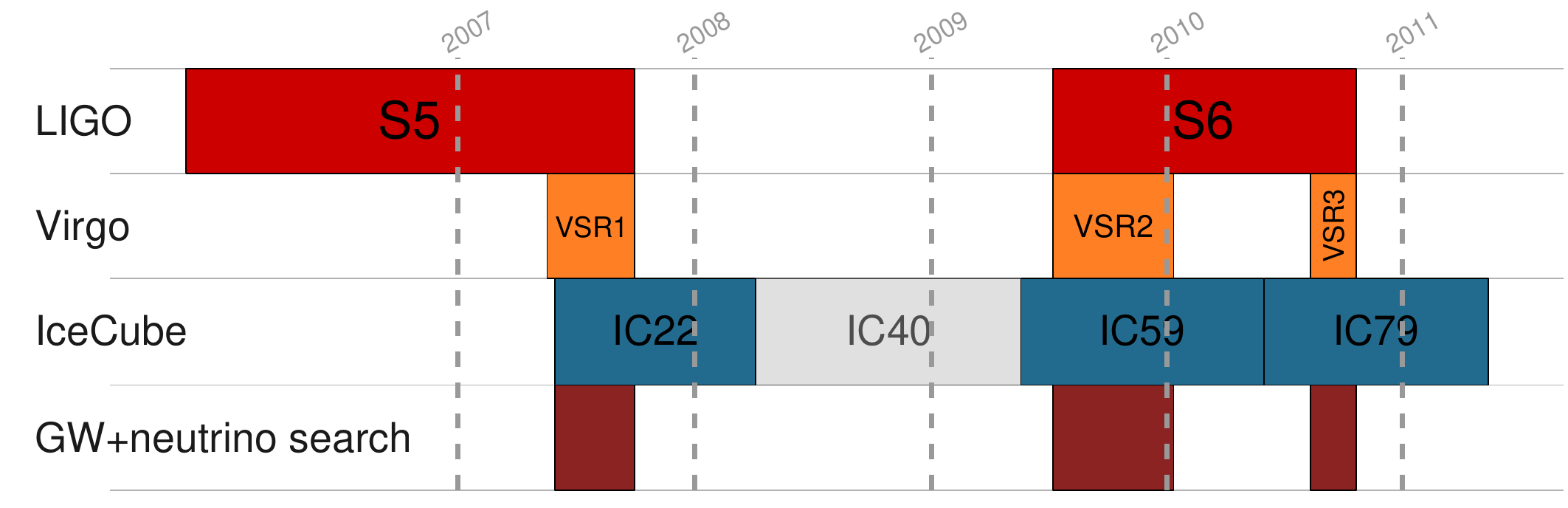}}
\end{center}
\caption{Observational periods of the initial LIGO and Virgo GW detectors, the partially completed IceCube detector, as well as the joint GW+neutrino search periods. The IC40 observation period was not coincident with LIGO-Virgo observation periods, and was therefore not included in this search. See also Section \ref{section:GWdata} and Table \ref{table:neutrinos} for more information on the GW and neutrino observation periods, respectively.}
\label{figure:observationalperiods}
\end{figure*}

The LIGO S5 observation period started on November 4, 2005 and ended on October 1, 2007, while S6 lasted from July 7, 2009 until October 20, 2010. Virgo VSR1 started on May 18, 2007 and ended on October 1, 2007. Virgo VSR2 ran from July 7, 2009 to January 11, 2010, while Virgo VSR3 ran from August 11, 2010 to October 20, 2010. The observational periods of LIGO and Virgo used in this analysis are shown graphically in Fig. \ref{figure:observationalperiods} in comparison to IceCube. The three GW search periods have similar noise spectra and have comparable sensitivities. The S6 period having a nominal factor of $\sim 2$ lower background noise level than the S5 period.

In the present analysis, we only use data when all available detectors were operational. This means that for the S5 observation period, data is analyzed only if H1, H2, L1 and V1 were operational, while for the S6 observation period, we require H1, L1 and V1 to be operational. The total duration of such data taking period for S5/VSR1, in coincidence with IceCube in its 22-strings configuration (IC22) is $\sim 64$\,days. There are $\sim 30$ days of coincident data between S6-VSR2 and IceCube in its 59-string configuration (IC59) and $\sim 26$ days of coincident data for S6-VSR3 and IceCube in its 79-string configuration (IC79).

We analyze GW data from the LIGO-Virgo observatories using the \emph{coherent WaveBurst} (cWB) analysis pipeline \cite{cWB0264-9381-25-11-114029}. The cWB pipeline searches for short GW transients with durations $\lesssim 1$\,s in the [64\,Hz, 2048\,Hz] frequency range, which includes the most sensitive frequency band of LIGO and Virgo ($\sim 150$\,Hz). The pipeline looks for ``generic" GW signals within this frequency range, with no further constraints on their waveform \cite{2005gr.qc.....8068K}. The cWB pipeline coherently combines information from all GW detectors to perform a constrained maximum likelihood analysis. The output of the pipeline is a list of astrophysical GW signal candidates, which we will refer to as GW events.

For each GW event, cWB records its (i) time of arrival, (ii) test statistic $\rho$ and (iii) directional distribution. The test statistic of a GW event is the coherent network amplitude $\rho$, which is proportional to the signal-to-noise ratio. The directional distribution is specified within $0.4^\circ\times0.4^\circ$ size pixels.

The significance of each event is used to calculate the GW false alarm rate (FAR$_{\rm gw}$), that is, the frequency of occurrence of events from background alone with a greater significance. In this analysis we consider only those GW events whose FAR$_{\rm gw}$, based on their significance, is $\leq1$\,day$^{-1}$. This FAR$_{\rm gw}$ was used for electromagnetic follow-up observations of GW events \cite{2012A&A...539A.124L}. The $\rho$ values corresponding to FAR$_{\rm gw}$\,$=1$\,day$^{-1}$ for the three observation periods are $\rho_{\rm s5/vsr1}\approx2.7$, $\rho_{\rm s6/vsr2}\approx3.3$, $\rho_{\rm s6/vsr3}\approx3.4$. Note that the different value for S5/VSR1 is due to the fact that the H2 detector is also used there.

\subsection{High-energy neutrino data}
\label{section:neutrino}

The IceCube Neutrino Observatory \cite{2006APh....26..155I}, located near the South Pole, is a cubic-kilometer Cherenkov detector optimized to detect neutrinos at the TeV-PeV energy scale. The full detector consists of 86 vertical strings, with a set of 60 digital optical modules (DOMs) on each string. These DOMs are used to detect Cherenkov light from neutrino-induced charged particles. The present search uses observational data from IceCube during its construction period, which can be characterized with the number of strings deployed at the time of the observation. Deployment was performed only during Austral summers, resulting in extended $\sim 1$\,yr observation periods using a constant number of deployed strings (see Fig. \ref{figure:observationalperiods}).

IceCube detects neutrinos through the Cherenkov photons from secondary charged particles. For the present search, the secondary particles are muons, which, at $>$\,GeV energies, travel long enough in ice before decaying for an accurate direction reconstruction. We apply the neutrino event selection of the standard IceCube point source analysis \cite{2008ICRC....5.1275Z,2008APh....29..299B,2009ApJ...701L..47A,2010ApJ...710..346A,2010APh....33..175B,2012Natur.484..351A,2014JInst...9P3009A}.  Due to the large flux of muons produced by cosmic ray interactions within the atmosphere, the search is constrained to muons moving upward through the detector. As the Earth is opaque to muons, atmospheric muons are filtered out in these directions. The remaining background is due to atmospheric neutrinos produced by cosmic ray interactions in the atmosphere, and which produce muons in the ice around or within the detector. The direction of the secondary muons are reconstructed to a precision of $\lesssim1^\circ$ in the TeV-PeV energy range \cite{IceCubeAhrens2004507}. At these energies, the difference between the direction of the incoming neutrino and the secondary muon is negligible.

Astrophysical neutrinos from individual sources can be identified by a localized excess in space and/or time above the atmospheric background. The energy spectrum of atmospheric neutrinos can also be used to differentiate between the background and astrophysical signals. Neutrinos produced in the atmosphere have a soft energy spectrum (e.g., \cite{2011PhRvD..83a2001A}) with a power law spectral index of $\sim -3.7$ above $100$\,GeV, compared to the harder astrophysical spectrum, with a spectral index of $\sim -2$ due to the expected Fermi acceleration of protons in the GRB outflow \cite{2006RPPh...69.2259M}.

The present search uses high-energy neutrino data from the IceCube Neutrino Observatory in its 22, 59 and 79-string configurations. Neutrino data was recorded with the 22-string configuration from May 31, 2007 to April 4, 2008. IceCube recorded data with its 59-string configuration from May 20, 2009 to May 30, 2010, and with its 79-string configuration from May 31, 2010 to May 12, 2011. The observational periods of IceCube used in this analysis are shown graphically in Fig. \ref{figure:observationalperiods} in comparison to the observation periods of LIGO and Virgo. The number of detected astrophysical neutrino candidates for the three observation periods, as well as their subsets that were detected during times when H1, L1 and V1 GW detectors were operational, are shown in Table \ref{table:neutrinos}.

\begin{table*}
\begin{tabular}{c|c|c|c|c|c}
  \hline
  \# of strings & Start & End & \# of $\nu$ & \# coincident with GW data & Ref. \\
  \hline
  22 & May 2007 & Apr 2008 & 4852   & 978 &  \cite{1538-4357-701-1-L47} \\
  59 & May 2009 & May 2010 & 43339  & 3363 &  \cite{0004-637X-779-2-132} \\
  79 & May 2010 & May 2011 & 50857  & 3892 &  \cite{0004-637X-779-2-132} \\
  \hline
\end{tabular}
\caption{Number of neutrinos for the three IceCube observation periods considered in the present analysis. First column (\# of strings): number of strings in the different IceCube configurations, corresponding to the three observation periods. Second column (Start): start date of observation period. Third column (End): end date of observation period. Fourth column (\# of neutrinos): number of neutrinos from the Northern hemisphere for the different observation periods. Fifth column (\# coincident with GW data): number of neutrinos that were detected during the time when H1, L1 and V1 were operational (for the first row, the operation of H2 was also required). Sixth column (Ref.): reference describing neutrino data selection.}
\label{table:neutrinos}
\end{table*}
For each neutrino, this analysis uses its (i) time of arrival, (ii) reconstructed direction, (iii) directional uncertainty and (iv) reconstructed energy. The present search uses the reconstructed neutrino energy as a test statistic to characterize the significance of the neutrino signal, while the reconstructed direction and its uncertainty are also used to determine the significance of the directional coincidence between neutrinos, GWs and galaxies.

Note that the energy measured by IceCube is necessarily a lower-limit on the actual energy of the neutrino. If a secondary muon is produced outside of the detector, an unknown fraction of the neutrino energy is transferred to the shower at the interaction vertex and the muon will lose energy before reaching the instrumented volume. These two effects are difficult to account for when reconstructing the neutrino's energy (c.f. \cite{2013NIMPA.703..190A,2014JInst...9P3009A}).

\subsection{Galaxy distribution}

The distribution of GW sources is expected to be highly non-uniform in the volume within a few tens of Mpc probed by initial LIGO and Virgo. We take advantage of this by looking for astrophysical sources in the directions of galaxies, and by assigning a weight to the direction of each galaxy proportional to the probability of detecting an astrophysical neutrino from that direction.  Here, we assume that common GW+neutrino sources are located within or near galaxies, and that the source population within/nearby a galaxy is proportional to the galaxy's blue luminosity (see \cite{2012PhRvD..85j3004B} and references therein). This estimate also assumes uniform directional sensitivity (IceCube's sensitivity weakly depends on the zenith angle; e.g., \cite{2011ApJ...732...18A}).

We take the list of galaxies with their locations and blue luminosities from the Gravitational-Wave Galaxy Catalog \cite{0264-9381-28-8-085016}. We assign each galaxy a weight that reflects the probability of detecting an astrophysical neutrino from the given galaxy.

Since we assume that the number of sources in a galaxy is proportional to $B$, the probability of detecting a neutrino from a source within this galaxy will also be proportional to $B$. The neutrino flux observed on Earth from any source within the galaxy source will scale with $r^{-2}$. Assuming that the typical number of expected neutrino events from such a flux from a source is $\ll 1$, the probability of detecting a single neutrino from a given galaxy is therefore proportional with $B/r^2$. This value will be our weight for the galaxy. Note that the factor would be $B^2/r^4$ for two neutrinos from two independent events from the same galaxy, but this factor would be very small and therefore we ignore it in this analysis.

We can also consider the detection of more than one neutrino from a given source. For this analysis we only consider up to two neutrinos, since three or more detected neutrinos from a single source would by itself be a detection if the neutrinos were all detected within a $500$\,s time window (see Section \ref{section:GWsignificance}). The probability of detecting a neutrino doublet from a single source, again with the assumption that the typical neutrino flux from a source is $\ll 1$, is proportional to $B/r^4$.

For simplicity, the analysis divides the sky into $0.1^\circ\times0.1^\circ$ pixels and sums the contribution from each galaxy to these pixels (see Fig. 1. in \cite{0264-9381-28-8-085016} and Fig. 5. in \cite{2012PhRvD..85j3004B}). We only take into account galaxies that are within the [5\,Mpc, 40\,Mpc] range. With the lower cutoff we only discard a small fraction of the potential events, while these events may produce a strong enough signal that could be detectable with GW or neutrino detectors independently. The farther cutoff takes into account that even in optimistic models the joint emission can only be detected out to $\lesssim 40$\,Mpc due to GW sensitivity.

\subsection{Analysis algorithm}
\label{section:analysisalgorithm}

The joint analysis aims to identify GW events and neutrinos that originate from a common astrophysical source \cite{2012PhRvD..85j3004B}. Below we refer to a temporally coincident GW event and one or more neutrinos as a GW+neutrino event. In this section we describe how the significance of a GW+neutrino event is calculated.

The joint analysis determines the significance of a GW+neutrino event by combining the significances and directional distributions of GW events, neutrinos and galaxies. The following measures are used:
\begin{enumerate}
\item GW signal test statistic: $\rho$;
\item neutrino energy: $E_\nu$;
\item probability of having more than 1 background neutrino temporally coincident with the GW (if applicable);
\item point spread functions of GW, neutrinos and galaxies;
\end{enumerate}
The p-values of these four measures are calculated separately, and then are combined into a joint p-value. Fig. \ref{figure:flowdiagram} illustrates how they are combined, while a detailed explanation is presented below. 

\begin{figure}
\begin{center}
\resizebox{0.49\textwidth}{!}{\includegraphics{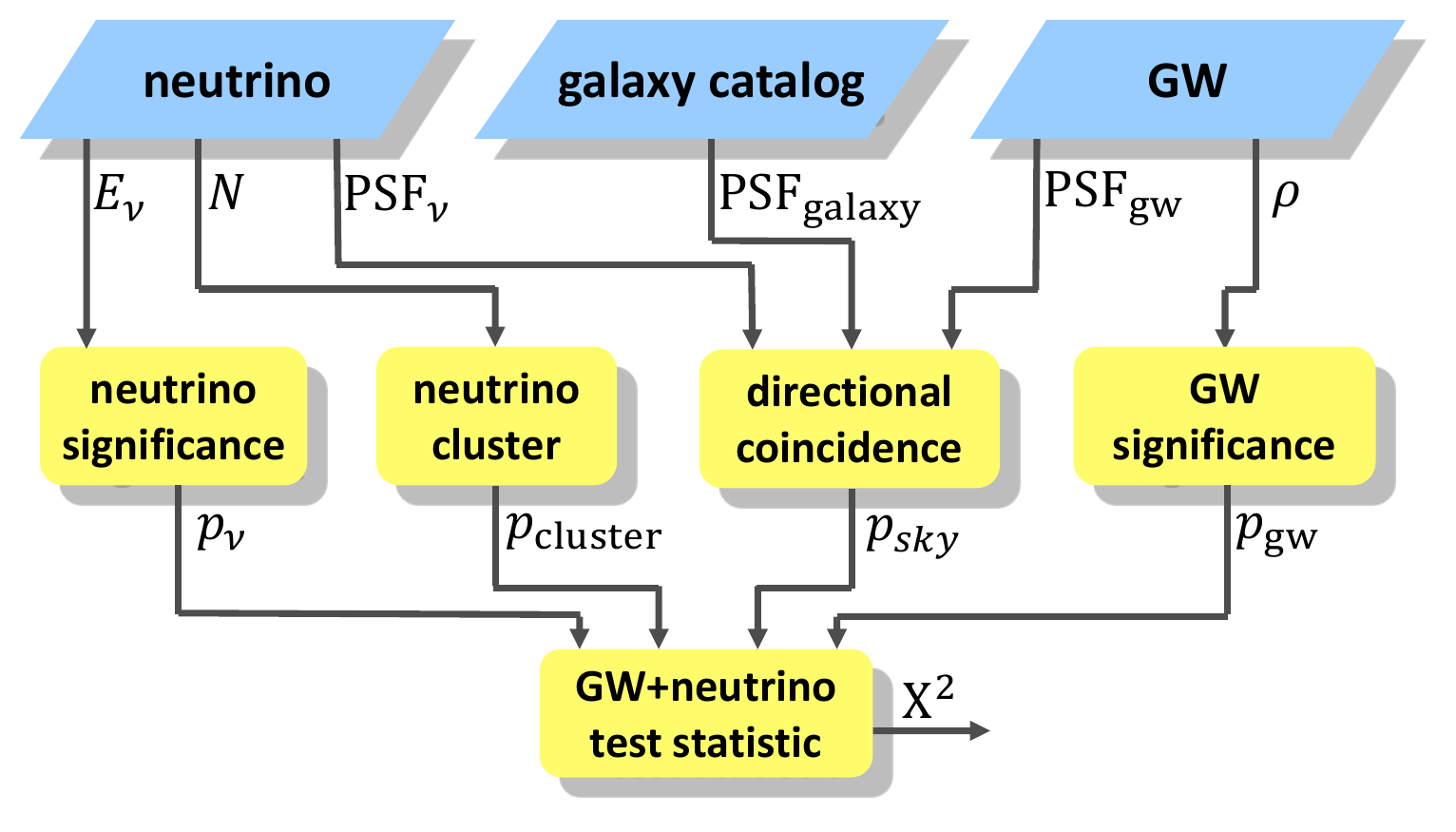}}
\end{center}
\caption{Flow diagram of the joint GW+neutrino analysis algorithm, showing how information on neutrinos, galaxies and GWs are combined into one test statistic. PSFs denote the point spread functions of GWs and neutrinos, and the weighted directional distribution of galaxies. See Section \ref{section:analysisalgorithm} for definitions of the parameters and for a detailed description of the algorithm.}
\label{figure:flowdiagram}
\end{figure}

\subsubsection{Neutrino significance}

We start with neutrinos detected by IceCube. We use the energy of a neutrino to calculate its significance; the neutrino's direction and the number of coincident neutrinos will be used separately (see Sections \ref{section:directionsignificance} and \ref{section:neutrinocluster}, respectively). For neutrino $i$ with reconstructed energy $E_\nu^{(i)}$, the p-value $p_{\nu}^{(i)}$ will be the probability of a background neutrino having energy $E_\nu^{\rm BG} \geq E_\nu^{(i)}$:
\begin{equation}
p_{\nu}^{(i)} = P(E_\nu^{\rm BG} \geq E_\nu^{(i)}).
\end{equation}
We conservatively assume here that the fraction of detected astrophysical neutrinos is negligible within all detected neutrinos, and use the energy estimators of all detected neutrinos for obtaining the p-values.

\subsubsection{Gravitational-wave significance}
\label{section:GWsignificance}

Next, we calculate the significance of GW event $j$ that is temporally coincident with neutrino $i$ above. The test statistic $\rho_j$ of GW event $j$ is obtained using the cWB pipeline. We first calculate the false alarm rate FAR$_{\rm gw}^{(j)}$ of the GW event by determining the rate at which background GW events occur with test statistic values $\rho_{\rm bg}\geq \rho_i$. We use this FAR$_{\rm gw}$ to calculate the p-value of GW event $j$, defined as the probability of finding a background GW event with $\rho \geq \rho_j$ that is temporally coincident with neutrino $i$:
\begin{equation}
p_{\rm gw}^{(i)}=1-\mbox{Pois}(0,2T\cdot\mbox{FAR}_{\rm gw}^{(j)}),
\end{equation}
where Pois$(k,\lambda)$ is the Poisson probability of $k$ outcomes with $\lambda$ average, and $T$ is the GW+neutrino coincidence time window.

We adopt an emission time window of $T=500$\,s following the recommendation of \cite{Baret20111}. We assume that a source emits both GWs and neutrinos over this 500\,s period, and consequently the maximum allowed time difference between a GW event and a neutrino is $500$\,s. For a detected neutrino, this corresponds to a total time window of $2T=1000$\,s ($\pm500$\,s) in which a GW event needs to be in order for the two events to be considered coincident. This $T$ time window is established based on the expected emission duration of long GRBs, being a conservative upper limit on the duration of long-GRB central engines' activity. It takes into account the observed duration of prompt gamma-ray emission, the duration of a relativistic jet burrowing through the stellar envelope, as well as potential precursor activity prior to prompt emission. The time window is likely also sufficient for other multimessenger transients, such as short GRBs or magnetar giant flares.

\subsubsection{Significance of neutrino cluster}
\label{section:neutrinocluster}

We further take into account that multiple neutrinos can be detected in coincidence with a GW event. Note that we do not search for multiple GW events from one astrophysical source. Two neutrinos are considered to have come from the same source only if they are temporally and directionally coincident. Neutrinos are considered temporally coincident if they arrive within $T$ from each other, following the same motivation as for the time window for GW events and neutrinos \cite{Baret20111}. We define directional coincidence as the existence of a common direction from which both neutrinos could originate from with probability $p>0.05$.

First, for a cluster of neutrinos, we calculate the significance of the neutrinos' test statistic by taking the product of the p-value of each neutrino within the cluster, to arrive at a significance
\begin{equation}
p_{\nu} = \prod_{\{k\}}p_{\nu}^{(k)},
\end{equation}
where the product is over all elements in the cluster.

Second, we assign a p-value to the coincidence of the cluster of $N$ neutrinos, defined as the probability that a total of $\geq N$ neutrinos occur within a $T$ interval. For background neutrinos detected with rate $f_\nu$ over the northern hemisphere and uniformly distributed in time, this probability is
\begin{equation}
p_{\rm cluster}(N) = 1 - \sum_{k=0}^{N-2} \mbox{Pois}(k,f_{\nu}\cdot T).
\end{equation}
Note that the sum goes to $N-2$ since we require at least 2 neutrinos for a cluster, i.e. $p_{\rm cluster}$ is a conditional probability, i.e. it is the probability of the detection of additional neutrinos given that the first neutrino has been detected. For only one neutrino, we have $p_{\rm cluster}(N=1)=1$.

\subsubsection{Significance of directional coincidence}
\label{section:directionsignificance}

We assign an additional significance to the directional coincidence of GWs, neutrinos and galaxies. We define the test statistic for directional coincidence to be the likelihood ratio of the GW event and neutrino coming from a common galaxy. We use this test statistic to calculate the p-value $p_{\rm sky}$ of directional coincidence. See Appendix A for the derivation of $p_{\rm sky}$ (see also \cite{2012PhRvD..85j3004B}).

\subsubsection{Combined significance}

After obtaining the p-values for GW events, neutrinos, as well as their directional coincidence, we combine these p-values using Fisher's method into one joint test statistic:
\begin{equation}
X^{2}_{i} = -2\ln\left[ p_{\rm sky}^{(i)}\,p_{\rm gw}^{(i)}\,p_{\rm cluster}(N)\,p_{\nu} \right]
\end{equation}
The background distribution of this combined test statistic $P_{\rm bg}(X^{2})$ is used to calculate the p-value for a given $X^{2}_i$:
\begin{equation}
p_{\rm gw\nu}^{(i)} = \int_{X^{2}_{i}}^{\infty}P_{\rm BG}(X^{2})\mbox{d}X^{2}.
\label{totalsignificance}
\end{equation}

It is often convenient to evaluate the significance of a GW+neutrino event using a joint false alarm rate (FAR) threshold. One can define a FAR corresponding to GW+neutrino event $i$ as $\mbox{FAR}_i = p_{\rm gw\nu}^{(i)}/t_{\rm obs}$, where $t_{\rm obs}$ is the duration of the joint observation period. This FAR means that the rate of background events identified by the joint analysis with $X^{2}\geq X^{2}_{i}$ will be FAR$_i$.

\subsection{Background distribution}
\label{section:background}

The background GW event level is generated by introducing time shifts between data from GW detectors. For a given realization, the time for each GW detector data is shifted compared to other detectors by a duration that is much greater than the travel time of a GW between the detectors, as well as the allowed GW transient duration. This ensures that, in the time-shifted data, no GW signal is coincident in the multiple detectors. For the analysis a total of 200 time shifts are produced, and cWB is used to search for GW events in these time-shifted data.

For a subset of background GW events identified by cWB, we assign neutrinos randomly selected from the list of neutrinos detected by IceCube. In practice this is implemented by changing the time of arrival of each neutrino to match the time of the corresponding GW background event. We keep the other parameters of the neutrino unchanged: the energy estimator and its uncertainty, as well as direction in local detector coordinates and its uncertainty.

We search for GW+neutrino events in the set of generated events using the joint analysis to obtain the background distribution of $X^2$. Since the  events are generated to have temporally coincident GW event and neutrino, we further need to account for the fact that a large fraction of the neutrinos is expected to have no temporally coincident GW event in the real data. This is accounted for using the GW event and neutrino rates and the observation time, and assuming uniform temporal distribution for both GW events and neutrinos.

\subsection{Simulated Signals}
\label{section:simulatedsignals}

We use simulated GW and neutrino signals to examine the sensitivity of the joint search to astrophysical events. For this analysis we adopt two different GW signal types. Both simulated GW signals have linearly polarized sine-Gaussian waveforms. This is a conservative choice as GW detectors are more sensitive to circularly polarized GWs (e.g., \cite{2010PhRvD..81j2001A}). We note that the observed GWs are circularly polarized for rotating sources viewed along their rotation axis, which is the case, e.g., for GRBs with highly collimated jets.

For linearly polarized GWs with polarizations $h_+$ and $h_\times$ of the GW strain, we can choose the coordinate system such that the signal's polarization is in the $+$ mode. Our simulated, linearly polarized GW signal can then be written as
\begin{eqnarray}
h_+(t) &=& h_0\sin(2\pi f_0 t)\exp[-(2\pi f_0 t)^2/2Q^2],\\
h_\times(t) &=& 0,
\label{equation:strain}
\end{eqnarray}
where $h_0$ is the maximum amplitude, $f_0$ is the characteristic frequency, and $Q/f_0$ is the characteristic duration of the GW signal. For simulated astrophysical GWs, two standard signal types with $f_0=153$\,Hz and $Q=8.9$, and $f_0=1053$\,Hz and $Q=9$. The first waveform has a characteristic frequency that is in the most sensitive band of the observatories, thus probing the reach of the measurement. For broadband emission such as compact binary mergers, this sensitive band will represent the most important contribution to detectability. The second waveform is at a higher frequency that may be more typical for some core collapse models and SGR flares. The value of $Q$ only weakly affects the search sensitivity, therefore the specific choice is not critical \cite{2008PhRvL.101u1102A}.

The two are standard gravitational waveforms that allow the results to be compared to other GW analyses (see, e.g., \cite{2012PhRvD..85l2007A}). The sensitivity of GW-only searches with respect to other waveforms can be seen, e.g., in \cite{2010PhRvD..81j2001A}. Simulated GWs were added to the recorded data with distinct strain amplitudes. The GW signal's root-sum-squared amplitude $h_{\rm rss}$ is defined as
\begin{equation}
h_{\rm rss}^2=\int (|h_+(t)|^2+|h_\times(t)|^2)dt.
\end{equation}
GWs are added with amplitudes in the range $h_{\rm rss}\in [10^{-22},2\times10^{-19}]$\,Hz$^{-1/2}$. For the standard-siren GW with $E_{\rm gw} = 10^{-2}$\,M$_\odot$c$^2$, this range corresponds to a source distance of $D\in [50\,\mbox{Mpc},50\,\mbox{kpc}]$. For each strain amplitude, we generate between $360 - 1114$ joint simulated events.

To generate a simulated GW signal, we first select a random direction on the sky from an isotropic distribution of sources with declination $\delta\geq0^\circ$, which defines the directional range in which IceCube is the most sensitive to TeV-PeV neutrinos. This direction determines the time-of-arrival difference between the detectors, as well as the antenna factors that define the projection of the strain amplitude into the data stream. Then a simulated GW signal with random polarization is added to the GW detector data.

Astrophysical high-energy neutrinos are simulated with a Monte Carlo simulation of the partially completed IceCube detector (see \cite{2009ApJ...701L..47A} for the IC22 simulation, and \cite{2013ApJ...779..132A} for IC59 and IC79). The simulation considers a neutrino arriving from the direction identical to the direction chosen above for the GW signal injection. The time of arrival of the neutrino and GW signal are chosen to be the same. Since we do not use any weight for the time difference of the GW event and the neutrino, choosing the same time or arrival is equivalent to selecting a random time for the GW event and the neutrino within time window $T$. The simulation performs the production of secondary muons from neutrino interactions in or in the vicinity of the detector, tracks them and their Cherenkov-light emission through the detector and generates the signals in the DOMs. Afterwards, the energy, direction and directional uncertainty of the neutrinos are reconstructed as with real data. The energies of the simulated neutrinos are drawn from a distribution $dN_\nu/dE_\nu\propto E_\nu^{-2}$ with an energy lower limit of $300\mbox{\,GeV}$, which takes into account the expected $E_\nu^{-2}$ scaling of Fermi acceleration, which is the acceleration mechanism typically considered for relativistic, shocked outflows (e.g., \cite{2007Ap&SS.309..119R}).

Each injected event is randomly assigned a host galaxy using the probability weights described above. The assigned galaxy is then treated as if its direction was identical to the direction of origin of the simulated GW and neutrino, i.e. it is ``placed" in the direction of the simulated signal. This treatment of the galaxy distribution increases the effective number of simulated astrophysical signals for a limited number of simulated GW signals. The obtained result is conservative: any non-uniformity in the galaxy distribution increases the advantage of using galaxies, and since the simulated source direction is uniformly distributed over the sky, it will, on average, decrease the significance of simulated signals.

\section{Results}
\label{section:results}

Our analysis has found no joint GW+neutrino event that was sufficiently significant to claim detection. We found 13, 56 and 69 temporally coincident GW events and neutrinos for the time periods spanning the S5/VSR1/IC22, S6/VSR2/IC59 and S6/VSR3/IC79 observation periods, respectively, which are consistent with the null hypothesis. We found no neutrino doublets that were coincident with a GW event, therefore in the following we focus on GW+neutrino events with a single neutrino.

The most significant GW+neutrino event identified by the analysis had $X^2=9.21$. It was detected on July 27, 2007 at 21:16:54 UTC, during the S5/VSR1/IC22 observation period. The reconstructed direction of the neutrino was right ascension 159.6$^\circ$ and declination 20.8$^\circ$. It hit 22 DOMs in IC22, making it a reconstructed event with an energy measure typical to atmospheric events. The GW event reached $\rho=4.11$. The p-value of the event, calculated for the full S5/VSR1/IC22 observation period, is $p=0.35$. The event was therefore not sufficiently significant to claim detection, and our results are consistent with the background-only null hypothesis. For comparison, we calculate here the $X^2$ value that would be necessary, on average for a GW+neutrino signal to reach a joint false alarm rate (FAR) of $10^{-2}$\,yr$^{-1}$ within the S5/VSR1/IC22 observation period, which approximately corresponds to 3$\sigma$ significance level. To obtain this $X^2$, we perform a Monte Carlo simulation as follows. For one realization, we select $N_{100}\equiv N_{\rm obs}\cdot(100\,\mbox{yr}/t_{\rm obs})$  simulated background GW+neutrino events (see Section \ref{section:background} and Fig. \ref{figure:CDF} (left)), where $N_{\rm obs}$ is the number of temporally coincident GW+neutrino events found during the S5/VSR1/IC22 observation period. For each realization, we select the simulated joint GW+neutrino background event with the maximum $X^2$ value. To obtain the expected $X^2$, we take the median maximum $X^2$ value over the realizations.  We find that an event in the S5/VSR1/IC22 observation period corresponding to a FAR of $10^{-2}$\,yr$^{-1}$ would need $X^2=14.3$. This FAR value is equivalent to a p-value of $p=\mbox{FAR} t_{\rm obs} = 1.7\times10^{-3}$.

A further examination of data around our most significant event revealed that the cause of the relatively large signal-to-noise ratio of this event was due to large signals present in the H1 and H2 detectors strain channels that were caused by an electromagnetic disturbance; this disturbance is clearly visible in magnetometer and voltage line monitor channel signals.

\begin{table}
\begin{tabular}{l|l}
  \hline
  Variable                & Definition \\
  \hline
  $h_{\rm rss}$           & observed GW root-sum-squared amplitude  \\
  $\rho$                  & GW test statistic \\
  $E_{\rm gw}^{\rm iso}$  & total emitted isotropic-equivalent GW energy\\
  $E_{\nu}^{\rm iso}$     & total emitted isotropic-equivalent $\nu$ energy\\
  $n_{\nu}^{\rm (ic86)}$  & average \# of detected $\nu$ by IC86 from 10\,Mpc*\\
  $X^2$                   & test statistic of GW+neutrino event \\
  $R_{\rm ul}$            & joint source rate upper limit \\
  \hline
\end{tabular}
\caption{List of variables and parameters used in the joint analysis, and their role. *detection with the full IceCube detector for a source in the northern hemisphere at 10\,Mpc distance.}
\label{table:variables}
\end{table}

To help the reader interpret our results, Table \ref{table:variables} includes the variables used describing source properties and the outcome of the search.

\subsection{Efficiency and Sensitivity}
\label{section:efficiency}

\begin{figure*}
\begin{center}
\resizebox{0.99\textwidth}{!}{\includegraphics{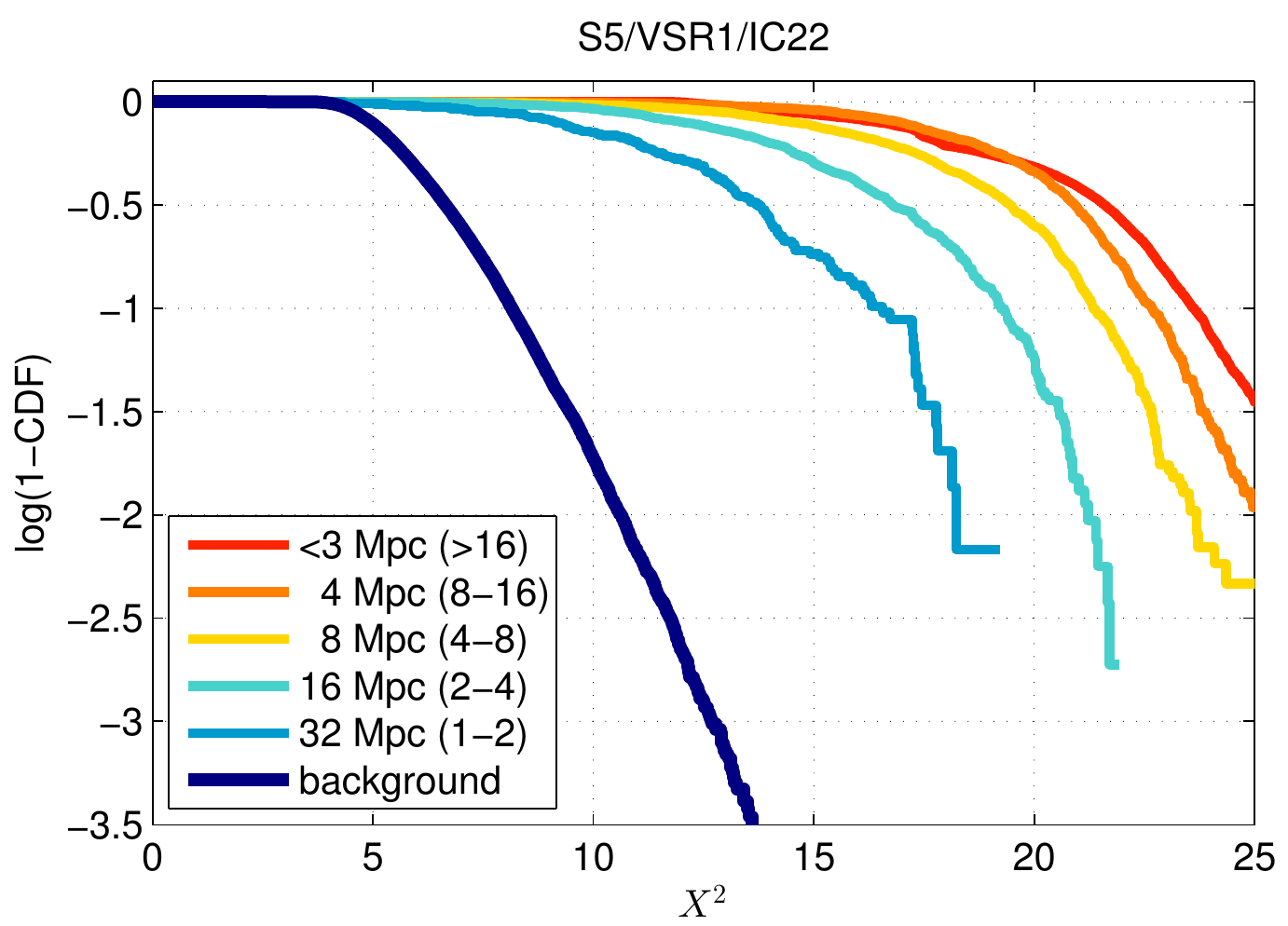}\includegraphics{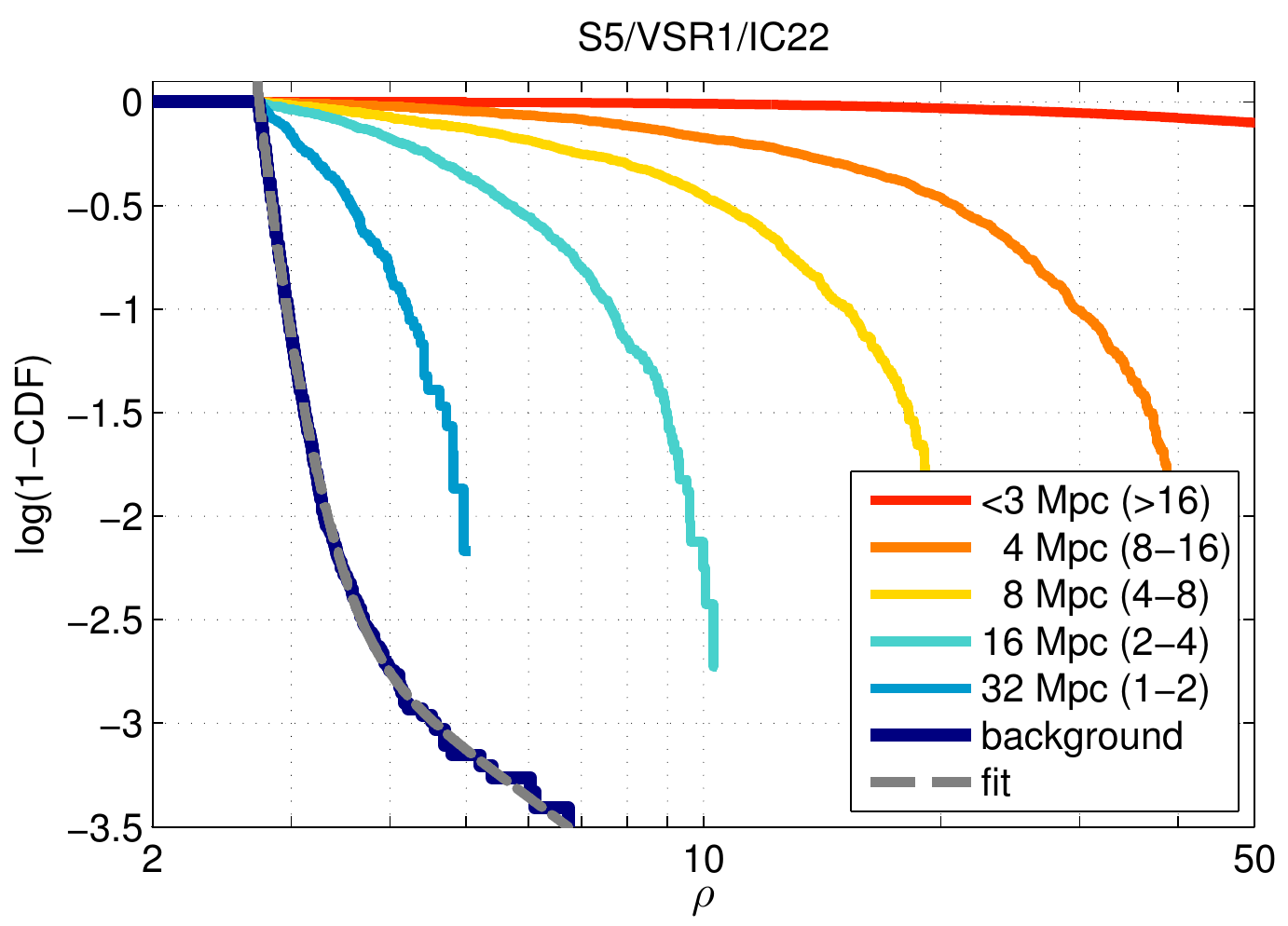}}
\end{center}
\caption{Complementary cumulative distribution function  (1 - CDF) of the test statistic of joint GW+neutrino (left) and GW-only (right) events for the background and for simulated astrophysical events, for the S5/VSR1/IC22 observation period. Results for simulated astrophysical events are shown for different characteristic distances, as indicated in the legend. The function 1-CDF grows with decreasing source distance. We used standard-siren GW injections ($E_{\rm gw}=10^{-2}$\,M$_{\odot}$c$^2$; sine-Gaussian waveform with characteristic frequency $f_0=\sim153$\,Hz and $Q=8.9$). For the GW+neutrino case, the results show events for which one neutrino was detected. We also express the characteristic distances with intervals of $h_{\rm rss}$ values, corresponding to the amplitude of a standard-siren GW signal at the characteristic distance. These $h_{\rm rss}$ intervals are shown in the legend in parentheses, in units of $10^{-22}$\,Hz$^{-1/2}$.  A fit to the background 1-CDF is also shown (see Section \ref{section:prospects}). The distributions for the S6/VSR2/IC59 and S6/VSR3/IC79 observation periods (not shown here) are qualitatively the same. The results show that the joint GW+neutrino search identifies simulated astrophysical signals with higher efficiency than the GW-only case search, and that a large fraction of the simulated signals can be clearly differentiated from the background.}
\label{figure:CDF}
\end{figure*}

We calculated the background distribution of the test statistic, $X^2$, using $10^5$ simulated background GW+neutrino events separately for each observation period, as well as the dependence of $X^2$ on the injected GW amplitude of a simulated astrophysical event separately for each observation period. Fig. \ref{figure:CDF} (left) shows the \emph{complementary cumulative distribution function} ($\equiv$\,1-cumulative distribution function; 1-CDF) of $X^2$ values obtained for the background, as well as for simulated joint astrophysical events for the S5/VSR1/IC22 observational period. For a given $X^{2}_{\rm i}$ value, 1-CDF indicates the fraction of events with $X^2$ values for which $X^2\geq X^{2}_{\rm i}$. Fig. \ref{figure:CDF} (right) shows a similar evaluation for the GW-only case, which uses the GW test statistic $\rho$ (for definition see \cite{cWB0264-9381-25-11-114029}). Comparing the two distributions, one can see that the joint analysis is more efficient at differentiating simulated astrophysical signals from the background. Results are qualitatively the same for the other two observation periods. While the sensitivity of the GW detectors is somewhat better for later runs, the general characteristics of the data are similar.

\begin{figure}
\begin{center}
\resizebox{0.49\textwidth}{!}{\includegraphics{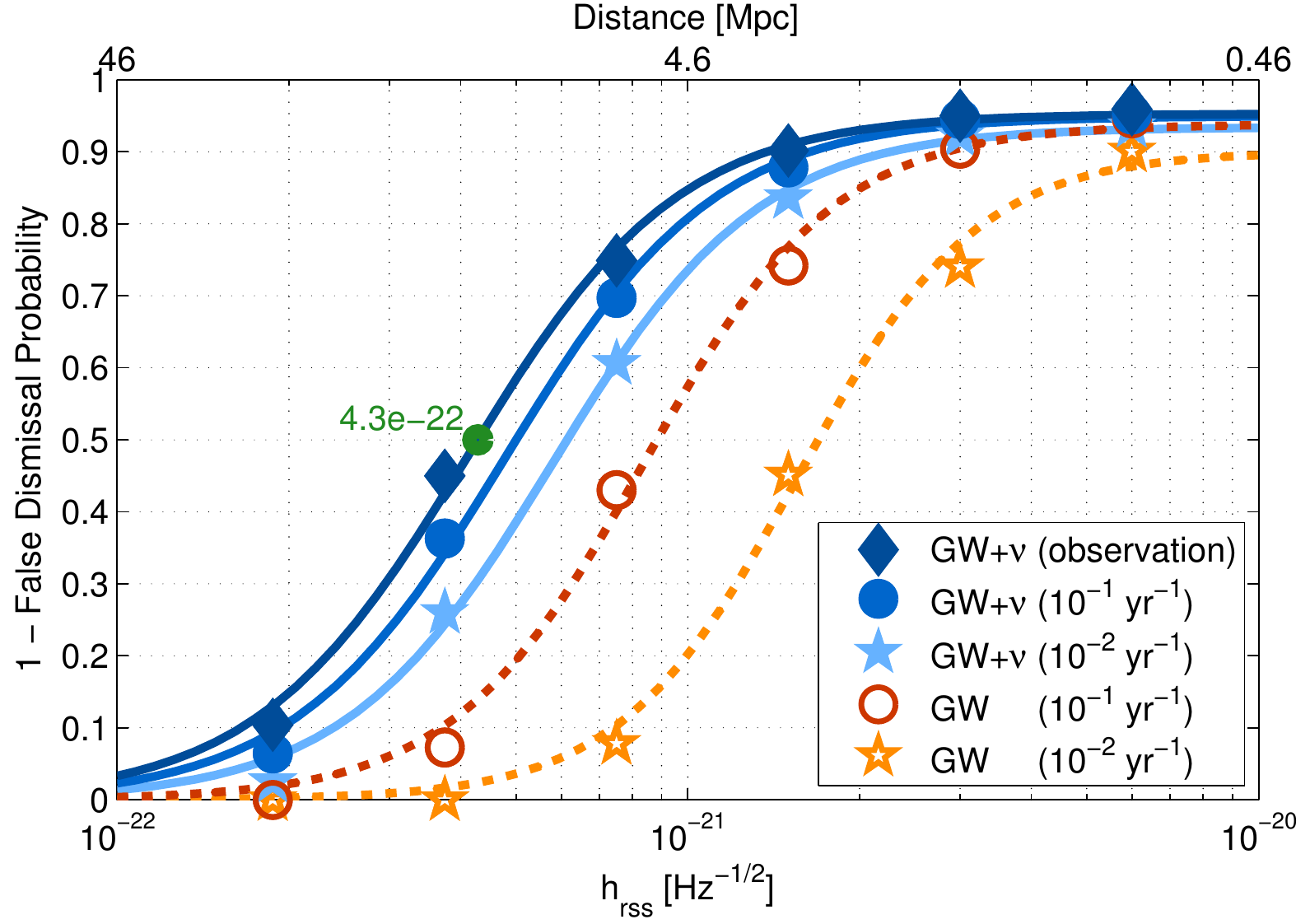}}
\end{center}
    \caption{Sensitivity of the joint analysis and the GW search as a function of GW $h_{\rm rss}$, for different FARs. The FAR values are shown in the legend. The threshold ``Observation" corresponds to the FAR of the most significant observed GW+neutrino event. For the simulated GW events, the standard-siren event was used (sine-Gaussian waveform; $f_0=153$\,Hz; $Q=8.9$). For the joint analysis, the efficiency is calculated given the detection of one neutrino from the source. For comparison, the top axis shows the source distances corresponding to the different $h_{\rm rss}$ for the standard-siren event. For the ``Observation" curve, we show $h_{\rm rss}^{50\%}=4.3\times10^{-22}$\,Hz$^{-1/2}$ (full circle). }
\label{figure:FARefficiency}
\end{figure}

We also evaluated the sensitivity of the joint search in detecting a simulated astrophysical GW+neutrino signal for different FAR thresholds. We determined the search sensitivity as a function of the GW root-sum-squared amplitude ($h_{\rm rss}$), and considered ``conditional sensitivity," given one neutrino has been detected from the source. The probability of detecting a neutrino from a given source is separately taken into account in, e.g., the upper limit estimates below. For a given FAR threshold and for a given $h_{\rm rss}$, we define sensitivity as the probability of a joint GW+neutrino event with 1 detected neutrino and with $h_{\rm rss}$ GW amplitude to be recovered by the search pipeline with a FAR below the FAR threshold. We used different thresholds: (i) the false alarm rate FAR$_{\rm max,obs}$, that corresponds to the highest test statistic value $X^2_{\rm max,obs}$,  measured during the observation period; (ii) FAR=0.1\,yr$^{-1}$ and (iii) FAR=0.01\,yr$^{-1}$ (see Fig. \ref{figure:FARefficiency}). We fit sigmoid curves onto the measured data to determine their trend. Not that the curves saturate somewhat below 1, which is due to the non-uniform directional sensitivity of the GW detector network, as well as uncertainties in the reconstructed direction.

Results for the S5/VSR1/IC22 observation period are shown in Fig. \ref{figure:FARefficiency}. We obtained similar results for the other two observation periods.

\subsection{Source rate upper limit}
\label{section:rateUL}

We use the sensitivity of the joint search, as well as of non-detection, to obtain astrophysical source rate upper limits. Here we build on and improve the analysis by Ref. \cite{PhysRevLett.107.251101}, that derived source rate upper limits using an effective detection volume defined by the distance at which sources can be detected with $50\%$ efficiency. We note that a previous GW+neutrino coincident search with data from initial LIGO-Virgo and the partially completed {\sc Antares} derived upper limits on the GW+neutrino source population using a different method \cite{2013JCAP...06..008A}. Ref. \cite{2013JCAP...06..008A} separately derives horizon distances for neutrino and GW signals (with $50\%$ and $90\%$ detection probabilities, respectively), and uses the smaller horizon distance of these two to define the effective search volume, which in turn determines the source rate upper limit. The upper limits derived in the LIGO-Virgo-{\sc Antares} analysis remained above the currently accepted source rates for long or short GRBs, as do the results presented here. Limits from LIGO-Virgo-{\sc Antares}, in principle, apply to a source population on the opposite (southern) hemisphere as the limits from the LIGO-Virgo-IceCube search (at $\sim $\,TeV neutrino energies).

We derive source rate upper limits using the sensitivity of the joint search as a function of $h_{\rm rss}$, determined for FAR$_{\rm max,obs}$ (see Section \ref{section:efficiency}). We calculate upper limits separately for the three observation periods, which are then combined into one overall upper limit.

We can express the sensitivity of the search, which is characterized above as a function of $h_{\rm rss}$, as a function of the source properties. The observed GW strain amplitude $h_{\rm rss}$ depends on the total isotropic-equivalent GW energy $E_{\rm gw}^{\rm iso}$ emitted by the source, as well as the source luminosity distance $D$:
\begin{equation}
h_{\rm rss} \approx \frac{kG^{1/2}}{\pi c^{3/2}}f_0^{-1}(E_{\rm gw}^{\rm iso})^{1/2}D^{-1},
\end{equation}
where $k$ is a constant that depends on the polarization of the emission, and $G$ is the gravitational constant. It is $k=(5/2)^{1/2}$ for circularly polarized emission that is expected from a rotating source whose rotation axis points towards the Earth, and $k=(15/8)^{1/2}$ for linearly polarized signals (e.g., \cite{ruleofthumb}). We conservatively adopt this latter value in the present analysis.

Let $\varepsilon_{\rm det}(E_{\rm gw}^{\rm iso},D)$ denote the probability that the joint search detects an astrophysical GW+neutrino event with test statistic $X^2>X^2_{\rm max,obs}$, given that one neutrino is detected from the source. This probability is shown as a function of $h_{\rm rss}(E_{\rm gw}^{\rm iso},D)$ in Fig. \ref{figure:FARefficiency}.

We now estimate the probability that a source will produce a GW+neutrino event with test statistic $X^2>X^2_{\rm max,obs}$. Let $N = n_{\nu}(D/10\,\mbox{Mpc})^{-2}$ be the expected number of neutrinos detected by IceCube from a source at luminosity distance $D$ in the Northern hemisphere, where $n_{\nu}$ is the expected number of detected neutrinos for a source located at 10\,Mpc. Note that $n_{\nu}$ depends on both the detector and the source. The probability that at least one neutrino will be detected from this source is
\begin{equation}
p(\geq1|D,n_{\nu}) = 1 - \mbox{Pois}(0,N).
\end{equation}
We further take into account the source's neutrino beaming factor $f_b$, which is expected to be similar to the gamma beaming factor. The beaming factor is defined such that $f_b^{-1}$ is the fraction of the full sky towards which neutrinos are emitted. We take $f_b=14$, which is a typical upper limit on the beaming factor of low-luminosity GRBs \cite{2007ApJ...662.1111L}. The probability that a randomly oriented source at distance $D$ will produce a GW+neutrino event that is detected with $X^2>X^2_{\rm max,obs}$ can be expressed as
\begin{equation}
P_{\rm det}(D,E_{\rm gw}^{\rm iso},n_{\nu}) =  f_b^{-1}p(\geq1|D,n_{\nu})\varepsilon_{\rm det}(E_{\rm gw}^{\rm iso},D).
\end{equation}
We point out that $P_{\rm det}$ is determined by the source distance and the source's GW ($E_{\rm gw}^{\rm iso}$) and neutrino ($n_{\nu}$) emission parameters. Note further that $P_{\rm det}$ is a probability for a joint event, while $p(\geq1|D,n_{\nu})$ is a probability for the neutrino-only case.

We place a limit on the rate of joint GW+neutrino events at $90\%$ confidence level. For this confidence level, we need to determine the source rate upper limit that corresponds to an average of $\geq2.3$ events over the measurement duration $t_{\rm obs}$ that would produce a GW+neutrino with $X^2>X^2_{\rm max,obs}$. Assuming a uniformly distributed source population, the rate density $R$ can be constrained below an upper limit $R_{\rm ul}$:
\begin{equation}
R_{\rm ul}(E_{\rm gw}^{\rm iso},n_{\nu}) = \frac{2.3}{t_{\rm obs}\int_0^{\infty}P_{\rm det}(D,E_{\rm gw}^{\rm iso},n_{\nu}) 4\pi D^2 \mbox{d}D}.
\end{equation}

\begin{figure}
\begin{center}
\resizebox{0.49\textwidth}{!}{\includegraphics{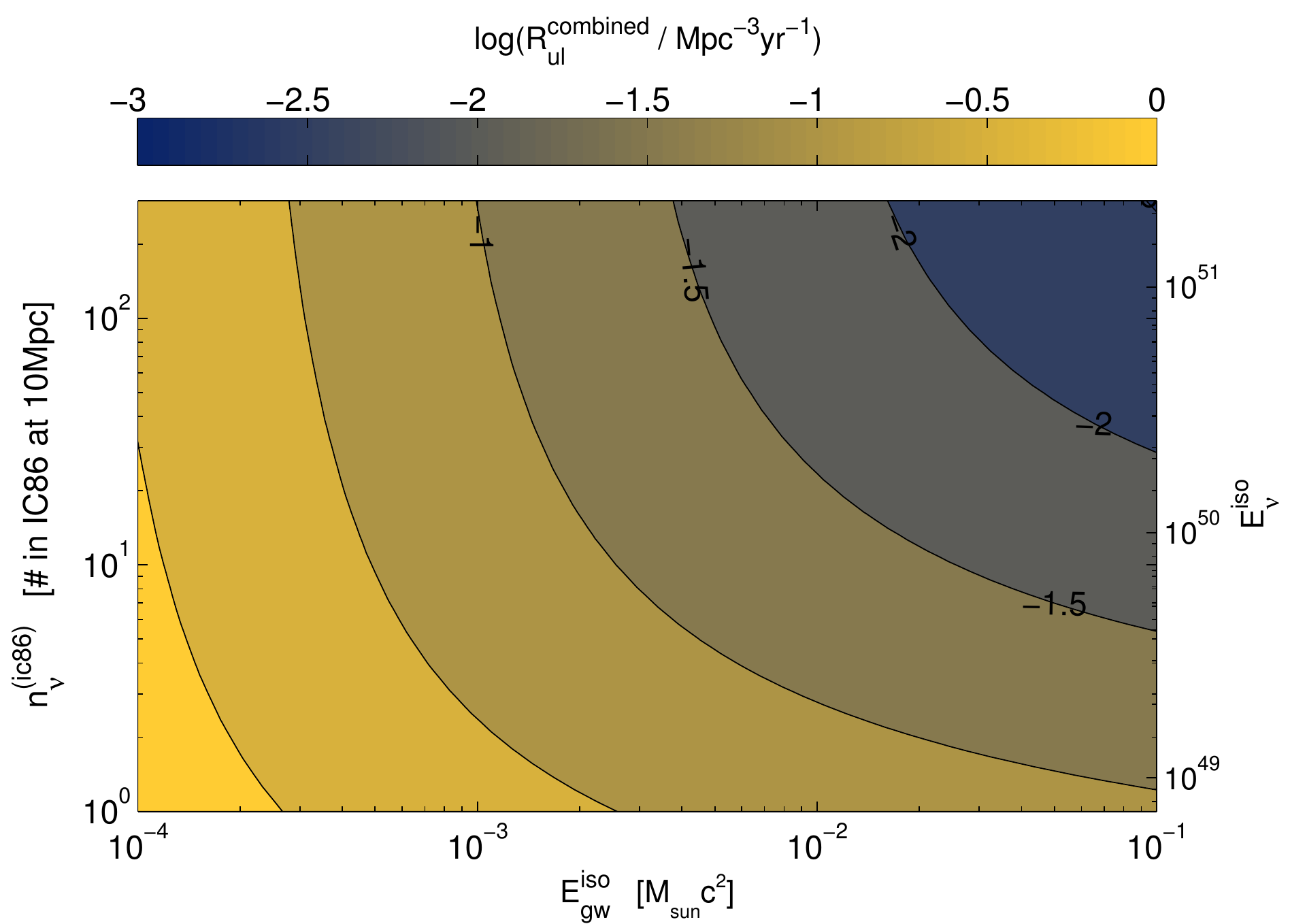}}
\end{center}
\caption{Joint GW+neutrino source rate upper limit of the present search as a function of isotropic-equivalent GW emission $E_{\rm gw}^{\rm iso}$ and neutrino emission. Neutrino emission is given both in terms of average number of neutrinos detected by IceCube with 86 strings from 10\,Mpc ($n_{\nu}^{\rm (ic86)}$), and in terms of emitted isotropic-equivalent neutrino energy ($E_{\nu}^{\rm iso}$; in all flavors). The results here assume an $E^{-2}$ neutrino energy spectrum (see Section \ref{section:neutrino}). The results shown here combine measurements from the three analyzed observation periods (S5/VSR1/IC22, S6/VSR2/IC59 and S6/VSR3/IC79).}
\label{figure:upperlimit}
\end{figure}

After calculating this source rate upper limit for the three observation periods, we combine these results to obtain one source rate upper limit for the whole search. Upper limits can be integrated by summing their inverse, such that
\begin{equation}
\frac{1}{R_{\rm ul}^{\rm combined}} = \sum_{i=1}^3 \frac{1}{R_{\rm ul}^{(i)}},
\end{equation}
where the sum is over the three observation periods. The resulting combined upper limit, $R_{\rm ul}^{\rm combined}$, is shown in Fig. \ref{figure:upperlimit} as a function of source emission parameters. For GW emission, the source emission strength is expressed with $E_{\rm gw}^{\rm iso}$. Characteristic GW energies range from $E_{\rm gw}^{\rm iso}\sim 10^{-2}$\,M$_\odot$c$^2$ for compact binary inspirals \cite{2009LRR....12....2S} to $E_{\rm gw}^{\rm iso}\lesssim 10^{-4}$\,M$_\odot$c$^2$ for core-collapse supernovae \cite{2009CQGra..26f3001O} (c.f., the horizontal axis of the Figure). For neutrino emission, the source emission strength is expressed as (i) isotropic-equivalent neutrino energy $E_{\nu}^{\rm iso}$ including all neutrino flavors, and (ii) as the average number $n_{\nu}^{\rm (ic86)}$ of neutrinos that would be detected from the source at 10\,Mpc distance, scaled to the full IceCube detector (with 86 strings). These two neutrino measures are connected as  $E_{\nu}^{\rm iso} = \kappa n_{\nu}^{\rm (ic86)}$, with $\kappa \approx 0.7\times 10^{49}$\,erg for an $E^{-2}$ neutrino spectrum \cite{PhysRevLett.107.251101}. To convert the sensitivity of the different partially completed IceCube runs to that of the full IceCube. We assume that sensitivity scales with the number of strings, which is a good approximation over the whole range of declinations searched; the relevant effective areas are presented in \cite{1538-4357-701-1-L47} and \cite{0004-637X-779-2-132}. Predicted characteristic neutrino emission energies include $E_{\nu}^{\rm iso} \sim 10^{51}$\,erg \cite{1997PhRvL..78.2292W}, and $E_{\nu}^{\rm iso} \sim 10^{50}$\,erg \cite{2005PhRvL..95f1103A,2008PhRvD..77f3007H} (see the secondary y-axis of Fig. \ref{figure:upperlimit}).

To put these results in context, we compare the obtained rate upper limits to those of potentially related astrophysical sources. One of the potential sources of interest is core-collapse supernovae with rapidly rotating cores, which may drive relativistic outflows that result in high-energy neutrino emission, and can also emit detectable GWs \cite{1998LRR.....1....8S,2000ApJ...542..453S,2002ApJ...565..430F,2003CQGra..20R.105A,2007PhRvD..75d4023B,2007CQGra..24S.171M,2010CQGra..27k4104C,2013CQGra..30l3001B}. The local ($z=0$) rate of core-collapse supernovae is $\sim 2 \times 10^{-4}$\,Mpc$^{-3}$yr$^{-1}$ \cite{2011ApJ...738..154H}, significantly below the upper limits that were obtained with the present search.

While the present results do not constrain current astrophysical models, we establish a baseline for future measurements by advanced GW detectors. The nominal sensitivity increase of advanced over initial GW detectors is a factor of $10$, which corresponds to a factor 1000 increase in sensitive volume. For the cases in which the GW component limits the joint sensitivity, this translates into a factor of 1000 improvement in the expected source rate upper limit upon no detection. For emission cases in which the neutrino component is limiting, this improvement will be smaller. For comparison, taking the standard-siren GW emission with $E_{\rm gw}^{\rm iso}=10^{-2}$\,M$_\odot$c$^2$, and a neutrino emission comparable to typical GRB gamma-ray emission, $E_{\nu}^{\rm iso}=10^{51}$\,erg \cite{1997PhRvL..78.2292W}, the upper limit obtained by the present search is $1.6 \times 10^{-2}$\,Mpc$^{-3}$yr$^{-1}$. In contrast, we project a source rate upper limit of $4\times10^{-4}$\,Mpc$^{-3}$yr$^{-1}$ for a 1-year observation period with advanced LIGO/Virgo and the full IceCube, which is comparable to the core-collapse supernova rate.

\subsection{Detection Sensitivity Improvement with the Joint Search}
\label{section:prospects}

An advantage of combining data from searches for GWs and neutrinos is the reduced total FAR. Requiring temporal and directional coincidence from a joint event are effective filters in reducing the background, beyond the identification capabilities of single-messenger searches. Low FAR searches are critical in identifying an astrophysical signal with high significance. This will be particularly important for the first GW discoveries. Below, we compare search sensitivities requiring low FAR for the joint GW+neutrino search and the GW-only and neutrino-only searches to characterize the improvement we can expect from the joint search.

To estimate the sensitivity of GW searches for low FAR, we take the S5/VSR1/IC22 observation period, which represents the longest duration of joint operation. To be able to estimate GW-only search sensitivities with low FAR, we extrapolate the distribution of the GW test statistic to increase the effective number of test-statistic samples. We fit the complementary cumulative distribution function of the GW test statistic $\rho$ (see Section \ref{section:GWdata}) with the empirical function $a_1(\rho-\rho_0^{(1)})^{\gamma_1}+a_2(\rho-\rho_0^{(2)})^{\gamma_2}$. The distribution of $\rho$, as well as the fit, are shown in Fig. \ref{figure:CDF}. With this empirical distribution we can estimate the average largest $\rho$ for a given measurement duration. This in turn can be used to calculate the expected $\rho$ threshold corresponding to a given FAR.

The efficiency of the joint search and a GW-only search in detecting a signal below a given FAR as a function of GW root-sum-squared strain amplitude $h_{\rm rss}$ is shown in Fig. \ref{figure:FARefficiency} for different FARs. The detection efficiency is given as a conditional probability, given 1 detected neutrino. The figure shows the difference between the efficiencies of the GW-only and the GW+neutrino searches. One can see that the sensitivity of the GW+neutrino search is much less dependent on the applied FAR threshold than the sensitivity of the GW-only search. The joint search is therefore increasingly advantageous for smaller FARs.

To quantitatively compare the sensitivity of the GW-only and GW+neutrino searches at low FAR, we consider the $h_{\rm rss}$ value for which the searches reach 50\,$\%$ efficiency. We denote the corresponding $h_{\rm rss}$ value with $h_{\rm rss}^{50\%}$. This is a meaningful comparison since $h_{\rm rss}^{50\%}$ is characteristic of the distance to which a GW search is sensitive to. The obtained efficiencies for different FAR levels are shown in Table \ref{table:sensitivity}. One can see that, for FAR\,$=0.01$\,yr$^{-1}$, the difference between the $h_{\rm rss}^{50\%}$ values of the GW-only and GW+neutrino searches is about a factor of $3$. Since the probed search volume scales with $(h_{\rm rss}^{50\%})^3$, the volume covered by the joint search increase by a factor of $\sim30$. Further, for a $5\sigma$ discovery, the difference between the GW-only and GW+neutrino searches is likely to grow even further. The joint search can therefore significantly expand the set of potentially detectable sources. The number of detected joint sources, nevertheless, will be affected by the probability of detecting a neutrino from a distant source as well as the beaming angle of the source.

\begin{table}
\begin{tabular}{l|c|c|c}
  \multicolumn{1}{c|}{Search} & FAR [yr$^{-1}$] & $h_{\rm rss}^{50\%}$ [Hz$^{-1/2}$] & Dist. [Mpc]\\
  \hline
  GW-only     & 0.1  & \,\,\,\,\,$9 \times 10^{-22}$  & 5 \\
  GW-only     & 0.01 & $1.7\times 10^{-21}$ & 3 \\
  GW+neutrino & 0.1  & \,\,\,\,\,$5\times 10^{-22}$  & 9 \\
  GW+neutrino & 0.01 & \,\,\,\,\,$6\times 10^{-22}$  & 8 \\
\end{tabular}
\caption{Distances and $h_{\rm rss}$ thresholds at which $50\%$ of astrophysical sources with standard-siren GW emission are detectable for different FAR values, for both the GW-only and the GW+neutrino searches. The GW+neutrino case assumes that 1 neutrino has been detected from the source.}
\label{table:sensitivity}
\end{table}

The above results also highlight the importance of GW detector characterization efforts. We see from Fig. \ref{figure:FARefficiency} that decreasing the FAR reduces the efficiency of the GW-only search much more significantly than that of the joint search, due to the non-Gaussian tail of the background GW event distribution. For a $5\sigma$ discovery over a 1-year observation period, assuming that the fit curve is valid for higher test statistic values, the required GW test statistic would need to be $\rho\approx 300$. This value is outside of the scale of the injections we performed, and shows that the non-Gaussian tail of the GW background distribution makes it difficult to raise the significance of a GW event. This result exemplifies that addressing the heavy-tail background distribution in GW detectors, e.g., in the context of detector characterization, will be critical for GW-only searches.

To compare the GW+neutrino search to the neutrino-only case, we emphasize that the joint search is in principle capable of detecting joint events with a single detected neutrino. A neutrino-only search, in the same time, would require at least 3 directionally and temporally coincident neutrinos for detection, without any electromagnetic or other counterpart (e.g., \cite{PhysRevLett.107.251101}). To compare the GW+neutrino and neutrino-only cases, we consider the difference between the characteristic distances from which 1 or 3 neutrinos are detected from a given source. For simplicity, we consider the case in which the respective detection probability is $50\%$. Let $D_{\nu,1}$ be the distance at which the probability of detecting at least one neutrino from a given source is $50\%$. This corresponds to an average of $\sim 0.7$ detected neutrinos. For the same source, detecting 3 or more neutrinos with $50\%$ probability corresponds to an average of $\sim 2.7$ detected neutrinos. The ratio between the characteristic distances of the neutrino-only search and the GW+neutrino search (given a GW trigger) is therefore
\begin{equation}
D_{\nu,3} / D_{\nu,1} \approx 2,
\end{equation}
corresponding to a search volume increase of a factor of $\sim 10$. Note that beaming will be the same for the neutrino-only and the GW+neutrino search, it will therefore not affect this difference. For the GW+neutrino case, a limitation will come from the finite distance from which GWs can be detected (up to hundreds of megaparsecs for advanced GW detectors, depending on the source's GW emission).

\subsection{PeV neutrinos}
\label{section:PeV}

Following IceCube's recent detection of $>100$\,TeV astrophysical neutrinos \cite{2013PhRvL.111b1103A,IceCubeScience}, it is beneficial to consider the extension of the present GW+neutrino search in this direction. While some of the $>100$\,TeV neutrinos have poor localization as they show a cascade-like hit-pattern in the detector, the present analysis pipeline can incorporate this limited direction reconstruction, as well as the temporal coincidence, to find joint GW+neutrino events.

We compared the time of arrival of the $>100$\,TeV neutrinos detected by IceCube to the time periods when LIGO-Virgo were operational. While some of the neutrinos were detected during the nominal LIGO-Virgo observation periods, in the present search we considered only times for which at least three detectors were operational. None of the $>100$\,TeV neutrinos had such triple coincidence. While this means that we presently did not determine meaningful limits on joint sources of GWs and $>100$\,TeV neutrinos, these searches can present an interesting direction for the advanced GW detector era, even in its early stages with limited GW direction reconstruction capabilities.

\section{Discussion and Outlook}
\label{section:discussion}

We searched for common transient sources of GWs and high-energy neutrinos using the initial LIGO and Virgo GW observatories (S5-6/VSR1-3 observation periods) along with the partially completed IceCube neutrino detector (22, 59, and 79 strings). We found no sufficiently significant joint event to claim detection. We used the results to derive a limit on the rate of common sources for a range of GW and neutrino emission parameters. Taking the standard-siren GW emission with $E_{\rm gw}^{\rm iso}=10^{-2}$\,M$_\odot$c$^2$ around $150$\,Hz and a neutrino emission energy $E_{\nu}^{\rm iso}=10^{51}$\,erg comparable to typical GRB gamma-ray emission, we constrain the source rate below $1.6\times10^{-2}$\,Mpc$^{-3}$yr$^{-1}$. While these results do not constrain current astrophysical models, we establish a baseline for future measurements by advanced GW detectors. We project a source rate upper limit of $4\times10^{-4}$\,Mpc$^{-3}$yr$^{-1}$ for a 1-year observation period with advanced LIGO/Virgo and the full IceCube, which is comparable to the core-collapse supernova rate.

We investigated the improvement of a GW+neutrino search over single-messenger searches with respect to (i) the obtained source rate upper limit and (ii) search sensitivity. We characterized the latter as the prospects of identifying an astrophysical signal with low FAR, which is necessary for discovery. For the first time we carried out a comparison with low FAR (1/100\,yr). We found that the GW+neutrino search performs significantly better than a GW-only search. Compared to the extrapolated GW background distribution, the horizon distance of the GW+neutrino search increased by a factor of $\sim 3$. This difference further increases with decreasing FAR. This demonstrates that the added information from multiple astronomical messengers will be especially important for high-significance detections. The difference for these detections is more pronounced than for source rate upper limits that allow higher FAR. Our results further indicate that the heavy-tail distribution of the GW-event test statistic limits the prospects of high-significance single-messenger detections, making GW detector characterization a critical step in increasing sensitivity for advanced GW detectors (e.g., \cite{2008CQGra..25r4004B,2010CQGra..27s4010C}). For characterizing the different sensitivities of GW+neutrino and neutrino-only searches, we find that the possibility of detection even with 1 observed neutrino for the joint analysis increases the search volume by a factor of $\sim 10$ compared to neutrino-only searches that require at least 3 coincident neutrinos for detection. This increased volume, nevertheless, may be limited by the range of GW detection.

The search algorithm developed for this analysis is readily usable with data from any number of advanced GW detectors and with neutrino observatories such as IceCube or KM3NeT. Observations may be especially interesting with early advanced GW detectors with only two operational observatories. The search can also be used to search for the highest-energy ($\sim$\,PeV) neutrinos coincident with GW signals, even with limited directional information.

With IceCube now completed and with the construction of advanced LIGO-Virgo finishing within a year, the next generation of multimessenger searches can commence soon, expanding our reach towards the high-energy, transient phenomena.

\begin{acknowledgments}

The authors gratefully acknowledge the support of the United States
National Science Foundation for the construction and operation of the
LIGO Laboratory, the Science and Technology Facilities Council of the
United Kingdom, the Max-Planck-Society, and the State of
Niedersachsen/Germany for support of the construction and operation of
the GEO600 detector, and the Italian Istituto Nazionale di Fisica
Nucleare and the French Centre National de la Recherche Scientifique
for the construction and operation of the Virgo detector. The authors
also gratefully acknowledge the support of the research by these
agencies and by the Australian Research Council,
the International Science Linkages program of the Commonwealth of Australia,
the Council of Scientific and Industrial Research of India,
the Istituto Nazionale di Fisica Nucleare of Italy,
the Spanish Ministerio de Econom\'ia y Competitividad,
the Conselleria d'Economia Hisenda i Innovaci\'o of the
Govern de les Illes Balears, the Foundation for Fundamental Research
on Matter supported by the Netherlands Organisation for Scientific Research,
the Polish Ministry of Science and Higher Education, the FOCUS
Programme of Foundation for Polish Science,
the Royal Society, the Scottish Funding Council, the
Scottish Universities Physics Alliance, The National Aeronautics and
Space Administration,
OTKA of Hungary,
the Lyon Institute of Origins (LIO),
the National Research Foundation of Korea,
Industry Canada and the Province of Ontario through the Ministry of Economic Development and Innovation,
the National Science and Engineering Research Council Canada,
the Carnegie Trust, the Leverhulme Trust, the
David and Lucile Packard Foundation, the Research Corporation, and
the Alfred P. Sloan Foundation. This article has LIGO document number LIGO-P1400046.

We acknowledge the support from the following agencies:
U.S. National Science Foundation-Office of Polar Programs,
U.S. National Science Foundation-Physics Division,
University of Wisconsin Alumni Research Foundation,
the Grid Laboratory Of Wisconsin (GLOW) grid infrastructure at the University of Wisconsin - Madison, the Open Science Grid (OSG) grid infrastructure;
U.S. Department of Energy, and National Energy Research Scientific Computing Center,
the Louisiana Optical Network Initiative (LONI) grid computing resources;
Natural Sciences and Engineering Research Council of Canada,
WestGrid and Compute/Calcul Canada;
Swedish Research Council,
Swedish Polar Research Secretariat,
Swedish National Infrastructure for Computing (SNIC),
and Knut and Alice Wallenberg Foundation, Sweden;
German Ministry for Education and Research (BMBF),
Deutsche Forschungsgemeinschaft (DFG),
Helmholtz Alliance for Astroparticle Physics (HAP),
Research Department of Plasmas with Complex Interactions (Bochum), Germany;
Fund for Scientific Research (FNRS-FWO),
FWO Odysseus programme,
Flanders Institute to encourage scientific and technological research in industry (IWT),
Belgian Federal Science Policy Office (Belspo);
University of Oxford, United Kingdom;
Marsden Fund, New Zealand;
Australian Research Council;
Japan Society for Promotion of Science (JSPS);
the Swiss National Science Foundation (SNSF), Switzerland;
National Research Foundation of Korea (NRF);
Danish National Research Foundation, Denmark (DNRF).

\end{acknowledgments}

\bibliographystyle{h-physrev}

\begin{thebibliography}{3}

\bibitem{LIGO0034-4885-72-7-076901}
B.~P. {Abbott} {\em et~al.},
\newblock Rep. Prog. Phys. {\bf 72}, 076901 (2009).

\bibitem{2003PhRvD..68h3001R}
S.~{Razzaque}, P.~{M{\'e}sz{\'a}ros}, and E.~{Waxman},
\newblock Phys. Rev. D {\bf 68}, 083001 (2003), arXiv:astro-ph/0303505.

\bibitem{PhysRevD.86.083007}
I.~Bartos, B.~Dasgupta, and S.~M\'arka,
\newblock Phys. Rev. D {\bf 86}, 083007 (2012).

\bibitem{2012Natur.484..351A}
R.~{Abbasi} {\em et~al.},
\newblock \nat {\bf 484}, 351 (2012), 1204.4219.

\bibitem{2013PhRvL.110x1101B}
I.~{Bartos}, A.~M. {Beloborodov}, K.~{Hurley}, and S.~{M{\'a}rka},
\newblock Physical Review Letters {\bf 110}, 241101 (2013), 1301.4232.

\bibitem{PhysRevLett.108.231101}
S.~H\"ummer, P.~Baerwald, and W.~Winter,
\newblock Phys. Rev. Lett. {\bf 108}, 231101 (2012).

\bibitem{2012ApJ...752...29H}
H.-N. {He} {\em et~al.},
\newblock \apj {\bf 752}, 29 (2012), 1204.0857.

\bibitem{2013RvMP...85.1401A}
S.~{Ando} {\em et~al.},
\newblock Reviews of Modern Physics {\bf 85}, 1401 (2013), 1203.5192.

\bibitem{2013CQGra..30l3001B}
I.~{Bartos}, P.~{Brady}, and S.~{M{\'a}rka},
\newblock Classical and Quantum Gravity {\bf 30}, 123001 (2013), 1212.2289.

\bibitem{0264-9381-28-11-114002}
T.~Accadia {\em et~al.},
\newblock Classical Quantum Gravity {\bf 28}, 114002 (2011).

\bibitem{2010CQGra..27h4006H}
G.~M. {Harry} and {LIGO Scientific Collaboration},
\newblock Classical and Quantum Gravity {\bf 27}, 084006 (2010).

\bibitem{AdV}
{The Virgo Collaboration},
\newblock Advanced virgo baseline design, 2009,
\newblock "VIR–027A–09".

\bibitem{2010CQGra..27h4003G}
H.~{Grote} and {the LIGO Scientific Collaboration},
\newblock Classical Quantum Gravity {\bf 27}, 084003 (2010).

\bibitem{2012CQGra..29l4007S}
K.~{Somiya},
\newblock Classical and Quantum Gravity {\bf 29}, 124007 (2012), 1111.7185.

\bibitem{LIGOindia}
B.~Iyer {\em et~al.},
\newblock LIGO-M1100296  (2011).

\bibitem{2013arXiv1304.0670L}
J.~{Aasi} {\em et~al.},
\newblock ArXiv e-prints  (2013), 1304.0670.

\bibitem{2006APh....26..155I}
{IceCube Collaboration} {\em et~al.},
\newblock Astroparticle Physics {\bf 26}, 155 (2006), astro-ph/0604450.

\bibitem{2009NIMPA.601..294A}
R.~{Abbasi} {\em et~al.},
\newblock Nuclear Instruments and Methods in Physics Research A {\bf 601}, 294
  (2009).

\bibitem{2010NIMPA.618..139A}
R.~{Abbasi} {\em et~al.},
\newblock Nuclear Instruments and Methods in Physics Research A {\bf 618}, 139
  (2010), 1002.2442.

\bibitem{ANTARES}
M.~{Ageron} {\em et~al.},
\newblock Nuclear Instruments and Methods in Physics Research A {\bf 656}, 11
  (2011), 1104.1607.

\bibitem{deJong2010445}
M.~de~Jong,
\newblock Nucl. Instrum. Methods Phys. Res., Sect. A {\bf 623}, 445  (2010).

\bibitem{Avrorin2011S13}
A.~Avrorin {\em et~al.},
\newblock {Nucl. Instrum. Methods Phys. Res., Sect. A} {\bf 626-627}, S13
  (2011).

\bibitem{2011A&A...535A.109A}
R.~{Abbasi} {\em et~al.},
\newblock \aap {\bf 535}, A109 (2011).

\bibitem{2003NIMPA.501..418F}
S.~{Fukuda} {\em et~al.},
\newblock Nuclear Instruments and Methods in Physics Research A {\bf 501}, 418
  (2003).

\bibitem{2002APh....16..361C}
L.~{Cadonati}, F.~P. {Calaprice}, and M.~C. {Chen},
\newblock Astroparticle Physics {\bf 16}, 361 (2002), hep-ph/0012082.

\bibitem{2008APh....28..516A}
N.~Y. {Agafonova} {\em et~al.},
\newblock Astroparticle Physics {\bf 28}, 516 (2008), 0710.0259.

\bibitem{2000NIMPA.449..172B}
J.~{Boger} {\em et~al.},
\newblock Nuclear Instruments and Methods in Physics Research A {\bf 449}, 172
  (2000), nucl-ex/9910016.

\bibitem{2013PhRvL.111b1103A}
M.~G. {Aartsen} {\em et~al.},
\newblock Physical Review Letters {\bf 111}, 021103 (2013).

\bibitem{IceCubeScience}
{IceCube Collaboration},
\newblock Science {\bf 342} (2013).

\bibitem{2013JCAP...06..030C}
I.~{Cholis} and D.~{Hooper},
\newblock \jcap {\bf 6}, 30 (2013), 1211.1974.

\bibitem{2013PhRvD..88h3007W}
W.~{Winter},
\newblock \prd {\bf 88}, 083007 (2013), 1307.2793.

\bibitem{2013PhRvL.111d1103K}
O.~E. {Kalashev}, A.~{Kusenko}, and W.~{Essey},
\newblock Physical Review Letters {\bf 111}, 041103 (2013), 1303.0300.

\bibitem{2013JCAP...01..028R}
E.~{Roulet}, G.~{Sigl}, A.~{van Vliet}, and S.~{Mollerach},
\newblock \jcap {\bf 1}, 28 (2013), 1209.4033.

\bibitem{2013arXiv1305.7404S}
F.~W. {Stecker},
\newblock ArXiv e-prints  (2013), 1305.7404.

\bibitem{2013PhRvD..87f3011H}
H.-N. {He}, T.~{Wang}, Y.-Z. {Fan}, S.-M. {Liu}, and D.-M. {Wei},
\newblock \prd {\bf 87}, 063011 (2013), 1303.1253.

\bibitem{2013PhRvD..88l1301M}
K.~{Murase}, M.~{Ahlers}, and B.~C. {Lacki},
\newblock \prd {\bf 88}, 121301 (2013), 1306.3417.

\bibitem{2014PhRvD..89h3003A}
L.~A. {Anchordoqui} {\em et~al.},
\newblock \prd {\bf 89}, 083003 (2014), 1306.5021.

\bibitem{2013arXiv1310.7194G}
M.~C. {Gonzalez-Garcia}, F.~{Halzen}, and V.~{Niro},
\newblock ArXiv e-prints  (2013), 1310.7194.

\bibitem{2013PhRvD..88h1302R}
S.~{Razzaque},
\newblock \prd {\bf 88}, 081302 (2013), 1309.2756.

\bibitem{2012PhRvL.108w1101H}
S.~{H{\"u}mmer}, P.~{Baerwald}, and W.~{Winter},
\newblock Physical Review Letters {\bf 108}, 231101 (2012), 1112.1076.

\bibitem{1997PhRvL..78.2292W}
E.~{Waxman} and J.~{Bahcall},
\newblock Physical Review Letters {\bf 78}, 2292 (1997), astro-ph/9701231.

\bibitem{2005PhRvL..95f1103A}
S.~{Ando} and J.~F. {Beacom},
\newblock Physical Review Letters {\bf 95}, 061103 (2005), astro-ph/0502521.

\bibitem{2006ApJ...651L...5M}
K.~{Murase}, K.~{Ioka}, S.~{Nagataki}, and T.~{Nakamura},
\newblock \apjl {\bf 651}, L5 (2006), astro-ph/0607104.

\bibitem{2003ApJ...589..861K}
S.~{Kobayashi} and P.~{M{\'e}sz{\'a}ros},
\newblock \apj {\bf 589}, 861 (2003), astro-ph/0210211.

\bibitem{2002ApJ...579L..63D}
M.~B. {Davies}, A.~{King}, S.~{Rosswog}, and G.~{Wynn},
\newblock \apjl {\bf 579}, L63 (2002), astro-ph/0204358.

\bibitem{2004PhRvD..69d4007V}
M.~H. {van Putten} {\em et~al.},
\newblock \prd {\bf 69}, 044007 (2004), gr-qc/0308016.

\bibitem{2007ApJ...658.1173P}
A.~L. {Piro} and E.~{Pfahl},
\newblock \apj {\bf 658}, 1173 (2007), astro-ph/0610696.

\bibitem{2011PhRvL.106y1102K}
K.~{Kiuchi}, M.~{Shibata}, P.~J. {Montero}, and J.~A. {Font},
\newblock Physical Review Letters {\bf 106}, 251102 (2011), 1105.5035.

\bibitem{2009ApJ...702.1171C}
A.~{Corsi} and P.~{M{\'e}sz{\'a}ros},
\newblock \apj {\bf 702}, 1171 (2009), 0907.2290.

\bibitem{2011LRR....14....1F}
C.~L. {Fryer} and K.~C.~B. {New},
\newblock Living Reviews in Relativity {\bf 14}, 1 (2011).

\bibitem{2006RPPh...69..971K}
K.~{Kotake}, K.~{Sato}, and K.~{Takahashi},
\newblock Reports on Progress in Physics {\bf 69}, 971 (2006),
  astro-ph/0509456.

\bibitem{2005PhRvD..71b4014S}
M.~{Shibata} and Y.-I. {Sekiguchi},
\newblock \prd {\bf 71}, 024014 (2005), astro-ph/0412243.

\bibitem{2005ApJ...633.1013I}
K.~{Ioka}, S.~{Razzaque}, S.~{Kobayashi}, and P.~{M{\'e}sz{\'a}ros},
\newblock \apj {\bf 633}, 1013 (2005), astro-ph/0503279.

\bibitem{2012PhRvD..85b4030Z}
B.~{Zink}, P.~D. {Lasky}, and K.~D. {Kokkotas},
\newblock \prd {\bf 85}, 024030 (2012), 1107.1689.

\bibitem{2013PhRvD..87j3008M}
D.~{Murphy} {\em et~al.},
\newblock \prd {\bf 87}, 103008 (2013), 1302.3915.

\bibitem{2000PhRvL..85.3761D}
T.~{Damour} and A.~{Vilenkin},
\newblock Physical Review Letters {\bf 85}, 3761 (2000), gr-qc/0004075.

\bibitem{2011PhRvD..84h5006B}
V.~{Berezinsky}, E.~{Sabancilar}, and A.~{Vilenkin},
\newblock \prd {\bf 84}, 085006 (2011), 1108.2509.

\bibitem{2006PhRvD..73j5001S}
X.~{Siemens} {\em et~al.},
\newblock \prd {\bf 73}, 105001 (2006), gr-qc/0603115.

\bibitem{2014PhRvL.112m1101A}
J.~{Aasi} {\em et~al.},
\newblock Physical Review Letters {\bf 112}, 131101 (2014), 1310.2384.

\bibitem{2013APh....43..134M}
P.~{M{\'e}sz{\'a}ros},
\newblock Astroparticle Physics {\bf 43}, 134 (2013), 1204.1897.

\bibitem{1993ApJ...405..273W}
S.~E. {Woosley},
\newblock \apj {\bf 405}, 273 (1993).

\bibitem{1999ApJ...524..262M}
A.~I. {MacFadyen} and S.~E. {Woosley},
\newblock \apj {\bf 524}, 262 (1999), astro-ph/9810274.

\bibitem{1992Natur.357..472U}
V.~V. {Usov},
\newblock \nat {\bf 357}, 472 (1992).

\bibitem{2006ARA&A..44..507W}
S.~E. {Woosley} and J.~S. {Bloom},
\newblock \araa {\bf 44}, 507 (2006), astro-ph/0609142.

\bibitem{1989Natur.340..126E}
D.~{Eichler}, M.~{Livio}, T.~{Piran}, and D.~N. {Schramm},
\newblock \nat {\bf 340}, 126 (1989).

\bibitem{1997A&A...317..140M}
E.~{Mueller} and H.-T. {Janka},
\newblock \aap {\bf 317}, 140 (1997).

\bibitem{2009CQGra..26f3001O}
C.~D. {Ott},
\newblock Classical and Quantum Gravity {\bf 26}, 063001 (2009), 0809.0695.

\bibitem{2013CRPhy..14..318K}
K.~{Kotake},
\newblock Comptes Rendus Physique {\bf 14}, 318 (2013), 1110.5107.

\bibitem{2013ApJ...778L..16H}
K.~{Hotokezaka} {\em et~al.},
\newblock \apjl {\bf 778}, L16 (2013), 1310.1623.

\bibitem{2011ApJ...732L...6R}
L.~{Rezzolla} {\em et~al.},
\newblock \apjl {\bf 732}, L6 (2011), 1101.4298.

\bibitem{2009LRR....12....2S}
B.~S. {Sathyaprakash} and B.~F. {Schutz},
\newblock Living Reviews in Relativity {\bf 12}, 2 (2009), 0903.0338.

\bibitem{1999CQGra..16R...1R}
F.~A. {Rasio} and S.~L. {Shapiro},
\newblock Classical and Quantum Gravity {\bf 16}, 1 (1999), gr-qc/9902019.

\bibitem{2005PhRvD..71h4021S}
M.~{Shibata}, K.~{Taniguchi}, and K.~{Ury{\= u}},
\newblock \prd {\bf 71}, 084021 (2005), gr-qc/0503119.

\bibitem{2010MNRAS.403..342H}
K.~{Hurley} {\em et~al.},
\newblock \mnras {\bf 403}, 342 (2010), 0907.2462.

\bibitem{2005PhRvL..95u1101O}
B.~J. {Owen},
\newblock Physical Review Letters {\bf 95}, 211101 (2005), astro-ph/0503399.

\bibitem{2006cosp...36..168W}
A.~{Watts},
\newblock {Neutron star oscillations and QPOs during magnetar flares},
\newblock in {\em 36th COSPAR Scientific Assembly}, , COSPAR Meeting Vol.~36,
  p. 168, 2006.

\bibitem{1992ApJ...392L...9D}
R.~C. {Duncan} and C.~{Thompson},
\newblock \apjl {\bf 392}, L9 (1992).

\bibitem{1995MNRAS.275..255T}
C.~{Thompson} and R.~C. {Duncan},
\newblock \mnras {\bf 275}, 255 (1995).

\bibitem{2005ApJ...634L..89G}
J.~D. {Gelfand} {\em et~al.},
\newblock \apjl {\bf 634}, L89 (2005), astro-ph/0503269.

\bibitem{2001PhRvL..87q1102M}
P.~{M{\'e}sz{\'a}ros} and E.~{Waxman},
\newblock Physical Review Letters {\bf 87}, 171102 (2001),
  arXiv:astro-ph/0103275.

\bibitem{2007arXiv0711.4791K}
H.~B.~J. {Koers} and R.~A.~M.~J. {Wijers},
\newblock ArXiv e-prints {\bf 711} (2007), 0711.4791.

\bibitem{2008PhRvD..77f3007H}
S.~{Horiuchi} and S.~{Ando},
\newblock \prd {\bf 77}, 063007 (2008), 0711.2580.

\bibitem{2007PhRvD..76h3009W}
X.-Y. {Wang}, S.~{Razzaque}, P.~{M{\'e}sz{\'a}ros}, and Z.-G. {Dai},
\newblock \prd {\bf 76}, 083009 (2007), 0705.0027.

\bibitem{2008CQGra..25k4039A}
Y.~{Aso} {\em et~al.},
\newblock Classical and Quantum Gravity {\bf 25}, 114039 (2008), 0711.0107.

\bibitem{2009NIMPA.602..268P}
T.~{Pradier},
\newblock Nuclear Instruments and Methods in Physics Research A {\bf 602}, 268
  (2009), 0807.2562.

\bibitem{PhysRevLett.107.251101}
I.~Bartos, C.~Finley, A.~Corsi, and S.~M\'arka,
\newblock \prl {\bf 107}, 251101 (2011).

\bibitem{2013JCAP...06..008A}
S.~{Adri{\'a}n-Mart{\'{\i}}nez} {\em et~al.},
\newblock \jcap {\bf 6}, 8 (2013), 1205.3018.

\bibitem{2007AAS...211.9903P}
J.~{Piscionere} {\em et~al.},
\newblock {LOOC UP: Locating and Observing Optical Counterparts to Unmodeled
  Pulses in Gravitational Waves},
\newblock in {\em American Astronomical Society Meeting Abstracts}, , Bulletin
  of the American Astronomical Society Vol.~38, pp. 910--+, 2007.

\bibitem{2008CQGra..25r4034K}
J.~{Kanner} {\em et~al.},
\newblock Classical Quantum Gravity {\bf 25}, 184034 (2008), 0803.0312.

\bibitem{2012A&A...539A..60A}
R.~{Abbasi} {\em et~al.},
\newblock \aap {\bf 539}, A60 (2012), 1111.7030.

\bibitem{2007APh....27..533K}
M.~{Kowalski} and A.~{Mohr},
\newblock Astroparticle Physics {\bf 27}, 533 (2007), astro-ph/0701618.

\bibitem{2012A&A...539A.124L}
{LIGO Scientific Collaboration} {\em et~al.},
\newblock \aap {\bf 539}, A124 (2012), 1109.3498.

\bibitem{2012A&A...541A.155A}
J.~{Abadie} {\em et~al.},
\newblock \aap {\bf 541}, A155 (2012), 1112.6005.

\bibitem{2012ApJS..203...28E}
P.~A. {Evans} {\em et~al.},
\newblock \apjs {\bf 203}, 28 (2012), 1205.1124.

\bibitem{2013arXiv1310.2314T}
{The LIGO Scientific Collaboration} {\em et~al.},
\newblock ArXiv e-prints  (2013), 1310.2314.

\bibitem{subthresholdGRB}
R.~Frey,
\newblock {Detecting GW bursts associated with sub-threshold Swift GRBs}, 2009,
\newblock (poster) GWDAW13.

\bibitem{subthresholdGRB2}
E.~D. Harstad, 2013,
\newblock Ph.D. Thesis.

\bibitem{2013arXiv1303.2174B}
L.~{Blackburn} {\em et~al.},
\newblock ArXiv e-prints  (2013), 1303.2174.

\bibitem{2004NJPh....6..114A}
P.~{Antonioli} {\em et~al.},
\newblock New Journal of Physics {\bf 6}, 114 (2004), astro-ph/0406214.

\bibitem{PhysRevLett.58.1490}
K.~Hirata {\em et~al.},
\newblock \prl {\bf 58}, 1490 (1987).

\bibitem{PhysRevLett.58.1494}
R.~M. Bionta {\em et~al.},
\newblock \prl {\bf 58}, 1494 (1987).

\bibitem{2013APh....45...56S}
M.~W.~E. {Smith} {\em et~al.},
\newblock Astroparticle Physics {\bf 45}, 56 (2013), 1211.5602.

\bibitem{2012PhRvD..85j3004B}
B.~{Baret} {\em et~al.},
\newblock \prd {\bf 85}, 103004 (2012), 1112.1140.

\bibitem{1992Sci...256..325A}
A.~{Abramovici} {\em et~al.},
\newblock Science {\bf 256}, 325 (1992).

\bibitem{2010arXiv1003.2481T}
{The LIGO Scientific Collaboration} {\em et~al.},
\newblock ArXiv e-prints  (2010), 1003.2481.

\bibitem{2012arXiv1203.2674T}
{The LIGO Scientific Collaboration} and {The Virgo Collaboration},
\newblock ArXiv e-prints  (2012), 1203.2674.

\bibitem{2014arXiv1403.6639T}
{The LIGO Scientific Collaboration} {\em et~al.},
\newblock ArXiv e-prints  (2014), 1403.6639.

\bibitem{2008PhRvL.101u1102A}
B.~{Abbott} {\em et~al.},
\newblock Physical Review Letters {\bf 101}, 211102 (2008), 0808.2050.

\bibitem{cWB0264-9381-25-11-114029}
S.~Klimenko, I.~Yakushin, A.~Mercer, and G.~Mitselmakher,
\newblock Classical Quantum Gravity {\bf 25}, 114029 (2008).

\bibitem{2005gr.qc.....8068K}
S.~{Klimenko}, S.~{Mohanty}, M.~{Rakhmanov}, G.~{Mitselmakher}, and {.},
\newblock ArXiv General Relativity and Quantum Cosmology e-prints  (2005),
  arXiv:gr-qc/0508068.

\bibitem{2008ICRC....5.1275Z}
J.-D.-D. {Zornoza} and D.~{Chirkin},
\newblock International Cosmic Ray Conference {\bf 5}, 1275 (2008).

\bibitem{2008APh....29..299B}
J.~{Braun} {\em et~al.},
\newblock Astroparticle Physics {\bf 29}, 299 (2008), 0801.1604.

\bibitem{2009ApJ...701L..47A}
R.~{Abbasi} {\em et~al.},
\newblock \apjl {\bf 701}, L47 (2009), 0905.2253.

\bibitem{2010ApJ...710..346A}
R.~{Abbasi} {\em et~al.},
\newblock \apj {\bf 710}, 346 (2010), 0907.2227.

\bibitem{2010APh....33..175B}
J.~{Braun} {\em et~al.},
\newblock Astroparticle Physics {\bf 33}, 175 (2010), 0912.1572.

\bibitem{2014JInst...9P3009A}
M.~G. {Aartsen} {\em et~al.},
\newblock Journal of Instrumentation {\bf 9}, 3009P (2014), 1311.4767.

\bibitem{IceCubeAhrens2004507}
J.~Ahrens {\em et~al.},
\newblock Astroparticle Physics {\bf 20}, 507  (2004).

\bibitem{2011PhRvD..83a2001A}
R.~{Abbasi} {\em et~al.},
\newblock \prd {\bf 83}, 012001 (2011), 1010.3980.

\bibitem{2006RPPh...69.2259M}
P.~{M{\'e}sz{\'a}ros},
\newblock Reports on Progress in Physics {\bf 69}, 2259 (2006),
  astro-ph/0605208.

\bibitem{1538-4357-701-1-L47}
R.~{Abbasi et al. (IceCube Collaboration)},
\newblock \apj Lett. {\bf 701}, L47 (2009).

\bibitem{0004-637X-779-2-132}
M.~G. Aartsen {\em et~al.},
\newblock The Astrophysical Journal {\bf 779}, 132 (2013).

\bibitem{2013NIMPA.703..190A}
R.~{Abbasi} {\em et~al.},
\newblock Nuclear Instruments and Methods in Physics Research A {\bf 703}, 190
  (2013).

\bibitem{2011ApJ...732...18A}
R.~{Abbasi} {\em et~al.},
\newblock \apj {\bf 732}, 18 (2011), 1012.2137.

\bibitem{0264-9381-28-8-085016}
D.~J. {White}, E.~J. {Daw}, and V.~S. {Dhillon},
\newblock CQG {\bf 28}, 085016 (2011).

\bibitem{Baret20111}
B.~{Baret} {\em et~al.},
\newblock Astropart. Phys. {\bf 35}, 1  (2011).

\bibitem{2010PhRvD..81j2001A}
J.~{Abadie} {\em et~al.},
\newblock \prd {\bf 81}, 102001 (2010), 1002.1036.

\bibitem{2012PhRvD..85l2007A}
J.~{Abadie} {\em et~al.},
\newblock \prd {\bf 85}, 122007 (2012), 1202.2788.

\bibitem{2013ApJ...779..132A}
M.~G. {Aartsen} {\em et~al.},
\newblock \apj {\bf 779}, 132 (2013).

\bibitem{2007Ap&SS.309..119R}
F.~M. {Rieger}, V.~{Bosch-Ramon}, and P.~{Duffy},
\newblock \apss {\bf 309}, 119 (2007), astro-ph/0610141.

\bibitem{ruleofthumb}
P.~J. {Sutton},
\newblock ArXiv e-prints  (2013), 1304.0210.

\bibitem{2007ApJ...662.1111L}
E.~{Liang}, B.~{Zhang}, F.~{Virgili}, and Z.~G. {Dai},
\newblock \apj {\bf 662}, 1111 (2007), astro-ph/0605200.

\bibitem{1998LRR.....1....8S}
N.~{Stergioulas},
\newblock Living Reviews in Relativity {\bf 1}, 8 (1998), arXiv:gr-qc/9805012.

\bibitem{2000ApJ...542..453S}
M.~{Shibata}, T.~W. {Baumgarte}, and S.~L. {Shapiro},
\newblock \apj {\bf 542}, 453 (2000), arXiv:astro-ph/0005378.

\bibitem{2002ApJ...565..430F}
C.~L. {Fryer}, D.~E. {Holz}, and S.~A. {Hughes},
\newblock Astrophysical Journal {\bf 565}, 430 (2002).

\bibitem{2003CQGra..20R.105A}
N.~{Andersson},
\newblock Classical and Quantum Gravity {\bf 20}, 105 (2003),
  arXiv:astro-ph/0211057.

\bibitem{2007PhRvD..75d4023B}
L.~{Baiotti}, R.~{de Pietri}, G.~M. {Manca}, and L.~{Rezzolla},
\newblock \prd {\bf 75}, 044023 (2007), arXiv:astro-ph/0609473.

\bibitem{2007CQGra..24S.171M}
G.~M. {Manca}, L.~{Baiotti}, R.~{DePietri}, and L.~{Rezzolla},
\newblock Classical and Quantum Gravity {\bf 24}, 171 (2007), 0705.1826.

\bibitem{2010CQGra..27k4104C}
G.~{Corvino}, L.~{Rezzolla}, S.~{Bernuzzi}, R.~{De Pietri}, and
  B.~{Giacomazzo},
\newblock Classical and Quantum Gravity {\bf 27}, 114104 (2010), 1001.5281.

\bibitem{2011ApJ...738..154H}
S.~{Horiuchi} {\em et~al.},
\newblock \apj {\bf 738}, 154 (2011), 1102.1977.

\bibitem{2008CQGra..25r4004B}
L.~{Blackburn} {\em et~al.},
\newblock Classical and Quantum Gravity {\bf 25}, 184004 (2008), 0804.0800.

\bibitem{2010CQGra..27s4010C}
N.~{Christensen}, {LIGO Scientific Collaboration}, and {Virgo Collaboration},
\newblock Classical and Quantum Gravity {\bf 27}, 194010 (2010).

\end{thebibliography}

\begin{appendix}
\hspace{1 mm}
\newline
\section{P-value of directional coincidence}

For the directional distributions of GWs, neutrinos and galaxies, a well-defined signal hypothesis exists, i.e., the joint source is located at a specific direction. Accordingly, for the directional distributions we define a likelihood ratio for each messenger that we combine into one joint likelihood ratio $\mathcal{L}$. For GWs, we take the point spread function derived by cWB \cite{cWB0264-9381-25-11-114029} as the signal likelihood $\mathcal{S}_{\rm gw}$. For neutrinos, we use a 2-dimensional Gaussian directional probability distribution function $\mathcal{S}_{\nu}$, with its center at the reconstructed neutrino direction, and standard deviation equal to the reconstructed directional uncertainty. For simplicity, we consider $\mathcal{S}_{\nu}$ to be 0 for directions at which it is $<10^{-4}$\,deg$^{-2}$ to ensure that very low-probability overlaps are not considered directionally coincident. For the galaxies we take the weighted galaxy distribution, as defined above, as the signal likelihood $\mathcal{S}_{gal}$. For simplicity, we use isotropic distribution for a background likelihood $\mathcal{B}$ for GWs, neutrinos and the galaxy distribution (the background neutrino distribution weakly depends on the zenith angle; e.g., \cite{2011ApJ...732...18A}). For joint GW+neutrino event $i$ we get the following joint likelihood ratio:
\begin{equation}
\mathcal{L}^{(i)}(\vec{x}_s) = \frac{\mathcal{S}_{\rm gw}^{(i)}(\vec{x}_s)\mathcal{S}_{gal}^{(i)}(\vec{x}_s)\prod_{\{j\}}\mathcal{S}_{\nu}^{(j)}(\vec{x}_s)}
{\mathcal{B}_{\rm gw}^{(i)}\mathcal{B}_{gal}^{(i)}\prod_{\{j\}}\mathcal{B}_{\nu}^{(j)}},
\end{equation}
where $\vec{x}_s$ is the unknown source direction, $\{j\}$  runs over the set of neutrinos in the cluster. Treating $\vec{x}_s$ as a nuisance parameter, we marginalize over the whole sky to arrive at a marginalized likelihood ratio
\begin{equation}
L^{(i)} = \int \mathcal{L}^{(i)}(\vec{x}_s) \mbox{d}\vec{x}_s.
\end{equation}
To determine the significance of a given marginalized likelihood ratio, we compare it to the distribution of $L$ for background data. Denoting this distribution with $P_{\rm BG}(L)$, we calculate its p-value $p_{\rm sky}$ that provides a measure of its significance:
\begin{equation}
p_{\rm sky}^{(i)}=\int_{L^{(i)}}^{\infty}P_{\rm BG}(L')\mbox{d}L'.
\end{equation}

\end{appendix}

\end{document}